\documentclass[aps,prd]{revtex4}
\usepackage{epsfig,epsf}
\usepackage{amsmath}
\usepackage{amsthm}
\usepackage{amsfonts}
\usepackage{amssymb}
\usepackage{dsfont}
\usepackage{multirow}
\usepackage{appendix}
\usepackage{slashed}
\usepackage[active]{srcltx}
\usepackage{psfrag}
\usepackage{subfigure}
\usepackage[colorlinks,citecolor=blue,urlcolor=red,linkcolor=red]{hyperref}

\begin{document}

\title{{\Large{\bf Semileptonic  and nonleptonic decays of $D$  into tensor
mesons with
light-cone sum rule
}}}

\author{S. Momeni}%
\email[]{e-mail: samira.momeni@phy.iut.ac.ir}
\author{R. Khosravi}%
\email[]{e-mail: rezakhosravi @ cc.iut.ac.ir}
\author{S. Ghaziasgar}%
\email[]{e-mail: sepide.ghaziasgar@ph.iut.ac.ir}

\affiliation{Department of Physics, Isfahan University of
Technology, Isfahan 84156-83111, Iran}

\begin{abstract}
Form factors of $D$  decays into   $J^{PC}=2^{++}$ tensor   mesons
are calculated in the light-cone sum rules  approach up to
twist-4 distribution amplitudes of the tensor meson.  The masses of
the tensor  mesons are comparable to that of the charm quark mass
$m_c$; therefore all terms including powers of $m_T/m_c$ are kept
out in the expansion of the two-particle distribution amplitude
$\langle T|\bar q_{1\alpha}(x) \, q_{2\delta}(0)|0\rangle$.
Branching ratios of the semileptonic $D \to T\,\mu
\,\bar{\nu}_{\mu}$ decays and nonleptonic $D\to T~ P ~(P=K, \pi)$
decays are taken into consideration. A comparison is also made
between our results and predictions of other methods and the
existing experimental values  for the nonleptonic case. The
semileptonic branching ratios are typically of the order of
$10^{-5}$, and the nonleptonic  ones show better agreement with the
experimental data in comparison to the Isgur-Scora-Grinstein-Wise predictions.
\end{abstract}

\maketitle

\section{Introduction}
Analysis of  heavy meson decays to the light ones, is a
useful tool to explore the CKM matrix  and $CP$ violations.
The $D$- meson decays  occurring by $c$ quark decay (in quark level) are
placed in the above-mentioned processes.

In the semileptonic decays, the form factors determine the
nonperturbative effects. The form factors of the semileptonic decays
of charmed meson $D_{(s)}$ to scalar, pseudoscalar or vector mesons
have been estimated by various approaches. In Refs.
\cite{Khodjamirian, BallD}, the light-cone sum rule (LCSR) approach has been used to study
the $D\to \pi(K)\,\ell\, \nu$ decays.
 The form factors of the nonleptonic $D\to \pi (K,
K^{*})\ell\,\nu$ transitions have been evaluated by the lattice QCD
 method in Ref. \cite{Abada, Aubin, Bernard}, while the
semileptonic processes $D\to \pi, \rho, K$ and $K^{*}$ have been
investigated by the heavy quark effective theory  in Ref.
\cite{WangD}. The semileptonic decays $D_{(s)} \to f_0
(K_0^*)\,\ell\, \nu$, $D_{(s)} \to \pi (K) \,\ell\, \nu$, and
$D_{(s)} \to K^* (\rho, \phi) \,\ell\, \nu$  have been studied in
the framework of the three-point QCD sum rules (3PSR)
\cite{Ignacio,Aliev, Ball2,Ball3,Ovchinnikov, Baier90, Dong,Mao}.
The $D$ meson decays into the axial vector meson, the $D_{q}\to
K_{1}\,\ell\,\nu\,(q=u, d, s)$ and $D\to\,a_{1},f_{1}(1285),
f_{1}(1420)$ transitions are analyzed by the 3PSR approach
\cite{Khosravi, Zuo}.

For the tensor meson, as the final state, the form factors have been
calculated in the  Isgur-Scora-Grinstein-Wise (ISGW) quark model and
its improved version, the ISGW2 model in Refs. \cite{Isgur1989,Scora1995}.
The observed $J^{PC}=2^{++}$ tensor mesons are: isovector meson
$a_{2}(1320)$, isodoublet state $K_{2}^{*}(1430)$, and isosinglet
mesons $f_{2}(1270)$ and $f_{2}'(1525)$. $a_{2}(1320)$ is a $(d
\bar{u})$ state, and $K_{2}(1430)$ is a $(s \bar{u})$ state, while the
wave functions of $f_{2}$ and $f_{2}'$ are defined as their mixing
angle:
\begin{eqnarray}\label{eq.1}
f_{2}(1270)=\frac{1}{\sqrt{2}}(u\bar{u}+d\bar{d})\cos\theta_{f_{2}}+s\bar{s}\sin\theta_{f_{2}},\nonumber\\
f_{2}'(1525)=\frac{1}{\sqrt{2}}(u\bar{u}+d\bar{d})\sin\theta_{f_{2}}-s\bar{s}\cos\theta_{f_{2}}.
\end{eqnarray}
Since $\pi\pi(KK)$  is the dominate decay of $f_{2}(f_{2}')$ (for
more information, see \cite{Hagiwara2002}), the mixing angle should
be small and it has been reported $\theta_{f_{2}}=7.8^{\circ}$
\cite{Li2001} and $\theta_{f_{2}}=(9\pm1)^{\circ}$
\cite{Hagiwara2002}. Therefore, $f_{2}(1270)$ is primarily a
$(u\bar{u}+d\bar{d})/\sqrt{2}$ state, while $f_{2}'(1525)$ is
dominantly $(s\bar{s})$ \cite{Cheng2010}.

In this paper  the form factors for the $D$ decays into light tensor
mesons ($T$)  in the LCSR approach are calculated. In this method,
the operator product is expanded  near the light cone, while the
nonperturbative hadronic matrix elements are parametrized by the
light-cone distribution amplitudes (LCDAs) of the tensor meson.

The paper is organized as follows: In Sec. \ref{se.2}, by using the
LCSR, the form factors of $D \to T \ell \bar{\nu}_{\ell}$ decays are
derived. In Sec. \ref{se.nu},  the numerical analysis of the LCSR
for the form factors is presented and the branching ratio values of
the  semileptonic and nonleptonic decays are evaluated. A comparison
is also made between  our results and the predictions of other
methods and experimental data in this section.

\section{$ D \to T$ form factors in the LCSR}\label{se.2}
In the LCSR method, to calculate  the $D \to T$ transition form factors, first,  the
correlation function
\begin{eqnarray}\label{eq.2}
\Pi_\mu (p', q) &=&i \int d^4x e^{iqx} \langle T (p', \lambda) |
{\cal{T}}  \{j_{\mu}^{int}(x) \,j_{D}^{\dag}(0) \}| 0 \rangle,
\end{eqnarray}
where $j_{D}=i \bar{q} \gamma_5 c$ is the interpolating current for
the ${D}$ meson and  $q=u, s$ and $d$ for $D=D^{0}, D^{+}_{s}$
and $D^{+}$, respectively, is considered. Moreover, in Eq. (\ref{eq.2}),
$j_{\mu}^{int}=\bar{q}' \gamma_\mu (1-\gamma_5) c$
is the interaction current in which   $q'=d$ for the $D^{0}\to a_{2}$
transition and $q'=s$ for $D^{0}(D_{s}^{+})\to K_{2}^{*}(f_{2}')$ decay.
In addition, $\sqrt{2}\,j_{\mu}^{int}(D^{+}\to f_{2})=j_{\mu}^{int}(D^{0}\to a_{2}^{*})$.

According to the general philosophy of the LCSR, the correlation
functions of Eq. (\ref{eq.2}) can be obtained in two ways: the
physical or phenomenological side and  the QCD or theoretical ones.
The form factors can be obtained by using the dispersion relation to
link these two parts.

Let us first consider the physical part of Eq. (\ref{eq.2}). By
inserting a complete set of hadrons with the same quantum numbers of
the $D$ meson between the currents and isolating the pole term of
the lowest $D$ meson, correlation function is obtained as
\begin{eqnarray}\label{eq.3}
\Pi_{\mu} (p',p)&=& \frac{\langle T (p',\lambda)|j_{\mu}^{int}(x)|D(p)\rangle \langle
D(p)|j_{D}^{\dag}(0)|0\rangle}{m^2_D-p^2}+\frac{1}{\pi}\int_{0}^{\infty}\frac{\rho_{\mu}^{h}(s)}{s-p^2}\,ds,
\end{eqnarray}
where, the first term in Eq. (\ref{eq.3})  represents the ground-state
$D$-meson contribution and the second term describes the
contributions of the higher states and continuum, while
$\rho_{\mu}^{h}$ is the spectral density for these states. These
spectral densities are approximated by evoking the quark-hadron
duality ansatz as
\begin{eqnarray}\label{eq.rho}
\rho_{\mu}^{h}(s)=\rho_{\mu}^{QCD}(s)\theta(s-s_{0}),
\end{eqnarray}
where $s_{0}$ is the continuum threshold chosen near the squared
mass of the lowest $D$-meson state. It follows from Eq. (\ref{eq.3})
that to  calculate  the form factors of the $B \to T$ transition
the matrix elements $\langle T
(p',\lambda)|j_{\mu}^{int}(x)|D(p)\rangle$ and $\langle
D(p)|j_{D}^{\dag}(0)|0\rangle$ are needed. The first matrix element
is defined in terms of the form factors as
\cite{Hatanaka2009, Hatanaka2010, Wang2011}
\begin{eqnarray}\label{eqffp}
\langle T (p',\lambda) |j_{\mu}^{int}(x)|D(p)\rangle &=&-i  {2 V
(q^2) \over m_{D}+ m_{T}} \epsilon_{\mu\nu\alpha\beta}\,
e^{*\nu}_{\lambda} p^\alpha p'^\beta - A_{1}(q^2)
e_\mu^{*\lambda}(m_{D} + m_{T})  \nonumber +
\frac{A_{2}(q^2)}{m_D+m_{T}} (e^{*\lambda}. q) (p+p')_\mu \\&+& 2
m_{T} \frac{( e^{*\lambda}.q)}{q^2} q_\mu [A_{3}(q^2) - A_{0}(q^2)],
\end{eqnarray}
with
\begin{eqnarray}\label{eqffc}
A_{3}(q^2)=\frac{m_D+m_{T}}{2m_{T}}A_{1}(q^2)-\frac{m_D-m_{T}}{2m_{T}}A_{2}(q^2)~~~   \rm{and} ~~~~~~
A_{3}(0)=A_{0}(0),
\end{eqnarray}
where $q=p-p'$,
$e_{\mu}^{\lambda}=\varepsilon_{\mu\nu}^{\lambda}p^{\nu}/m_{D}$.
Moreover, In Eq. (\ref{eqffp}), $V, A_{i}(i=0,..,3)$ are the form
factors of  $D \to T$ transition.

For simplicity, the following definitions are used:
\begin{eqnarray}\label{ffdef}
\mathcal{V} (q^2)&=&-{2 V (q^2) \over m_{D}+ m_{T}},~~~~~~~
\mathcal{A}_{1}(q^2)=-A_{1}(q^2)(m_{D} + m_{T}),  \nonumber\\
\mathcal{A}_{2}(q^2)&=&\frac{A_{2}(q^2)}{m_D+m_{T}},~~~~~~~~~
\mathcal{A}_{3}(q^2)= \frac{2m_{T}}{q^2} [A_{3}(q^2) - A_{0}(q^2)].
\end{eqnarray}
On the other hand, the second matrix element in Eq. (\ref{eq.3}) is
defined in  terms of the  $ D$-meson leptonic decay constant and
mass as
\begin{eqnarray}\label{sdef}
\langle
D(p)|j_{D}^{\dag}(0)|0\rangle=\frac{f_{D}m_{D}^{2}}{m_{c}+m_{q}}.
\end{eqnarray}
Using Eqs. (\ref{eq.rho}), (\ref{eqffp}), (\ref{ffdef}) and (\ref{sdef}),
these hadronic representation can be obtained for $\Pi_{\mu}(p,
p')$:
\begin{eqnarray}\label{hadrep}
\Pi_{\mu} (p',p)&=&
\frac{f_{D}m_{D}}{m_{c}+m_{q}}\frac{1}{m^2_D-p^2}
\Bigg\{i\,\mathcal{V} (q^2) \epsilon_{\mu\nu\alpha\beta}\,
\varepsilon^{*\nu\rho}_{\lambda} p^\alpha p'^\beta p_{\rho} +
\mathcal{A}_{1}(q^2)  \varepsilon_{\mu\sigma}^{*\lambda}p^{\sigma}
\nonumber +\mathcal{A}_{2}(q^2) \varepsilon^{*\lambda}_{\alpha\beta}
q^{\alpha} p^{\beta} (p+p')_\mu \\&+&\mathcal{A}_{3}(q^2)
\varepsilon^{*\lambda}_{\alpha\beta} q^{\alpha} p^{\beta} q_\mu
\Bigg\}+\frac{1}{\pi}\int_{s_{0}}^{\infty}\frac{\rho_{\mu}^{QCD}(s)}{s-p^2}\,ds.
\end{eqnarray}
To obtain the theoretical part of Eq. (\ref{eq.2}) in the LCSR approach,
the $\mathcal{T}$ product of currents should be expanded near the
light cone $x^{2}\simeq 0$. After contracting the $c$ quark field,
\begin{eqnarray}\label{eqth1}
\Pi_\mu (p', q) &=&-\int d^4x e^{iqx} \langle T (p', \lambda) |
\bar{q}'(x) \gamma_\mu (1-\gamma_5)\,S^{c}(x, 0)\, \gamma_5 \, \bar{q}(0)
| 0 \rangle,
\end{eqnarray}
where $S^{c}(x, 0)$ is the full propagator of the $c$
quark, is obtained. In this paper, just the free propagator is considered as
\begin{eqnarray}\label{prop}
S^{c}(x, 0)=\int \frac{d^4k}{(2\pi)^4} e^{-ikx} \frac{\not\!k +
m_c}{k^2-m_c^2}.
\end{eqnarray}
Using the Fierz rearrangement formula in Eq. (\ref{eqth1}), it
follows that in order to calculate the theoretical part, the matrix
elements of the nonlocal operators between $T$-meson and vacuum
states are needed. Two-particle distribution amplitudes for the
tensor meson $T$ are  given in \cite{Yangt2011, Chengt2010},
\begin{eqnarray} \label{eqdas}
&&\langle T(p',\lambda)|\bar q_{1\alpha}(x) \,
q_{2\delta}(0)|0\rangle = -\frac{i}{4} \, \int_0^1 du \,  e^{i u  p'
x}   \Bigg\{ f_T m_T^2 \Bigg[    \not\!p' \,
\frac{\varepsilon^{*\lambda}_{\mu\nu} x^\mu x^\nu}{(p'x)^2} \,
\Phi^T_\parallel(u) -\not\!x{\varepsilon^{*\lambda}_{\mu\nu}x^\mu x^\nu \over 2(p'x)^3} m_T^2 \bar{g}_{3}^{T}(u)  \nonumber \\
&& \qquad +  \left({\epsilon^{*\lambda}_{\mu\nu}x^\nu \over
p'x}-p'_\mu{\varepsilon^{*\lambda}_{\nu\beta}x^\nu x^\beta\over
(p'x)^2}\right)\gamma^\mu\, g_v^{T}(u) +  \frac{1}{2}
\epsilon_{\mu\nu\rho\sigma}\gamma^\mu
\varepsilon^{*\nu\beta}_{\lambda} x_\beta p'^\rho x^\sigma\gamma_5{1\over p'x}\, g_a^{T}(u)  \Bigg] \nonumber \\
&&\qquad - \,{i\over 2}f^{\perp}_T m_T \Bigg[ \frac{\sigma^{\mu\nu}
\left(\varepsilon^{*\lambda}_{\mu\beta} x^\beta p'_\nu -
\varepsilon^{*\lambda}_{\nu\beta} x^\beta p'_\mu\right)}{p'x}
\Phi^T_\perp(u) + \frac{\sigma^{\mu\nu}(p'_\mu x_\nu - p'_\nu x_\mu)
m_T^2\varepsilon^{*\lambda}_{\rho\beta}x^\rho x^\beta}{(p'x)^3} \,\bar h_t^{T}(u) \nonumber  \\
&& \qquad + \sigma^{\mu\nu}\left(\varepsilon^{*\lambda}_{\mu\beta}
x^\beta x_\nu - \varepsilon^{*\lambda}_{\nu\beta} x^\beta
x_\mu\right) \frac{m_T^2}{2(p'x)^2}\, \bar{h}_{3}^T(u)+
\varepsilon^{*\lambda}_{\mu\nu}x^\mu x^\nu {m_T^2\over p'x}\,h_s^{T}
(u)\Bigg]+{\cal O}(x^2)\Bigg\}_{\delta\alpha}\,,
\end{eqnarray}
where $x^2 \neq 0$ and
\begin{eqnarray}
\bar{g}_{3}(u) &=&{g}_{3}(u)+\Phi_\parallel (u)-2 g_v(u),~~~~~
 \bar h_t(u) = h_t(u) - \frac{1}{2} \Phi_\perp(u),~~~~~
\bar{h}_{3}(u) ={h}_{3}(u)-\Phi_\perp(u).~~~~~
\end{eqnarray}
In Eq. (\ref{eqdas}), $\Phi_{\parallel} $ and $\Phi_{\perp}$ are
the twist-2 functions; $g_{v}, g_a, h_{t}$ and $ h_s$  are the twist-3 functions;   ${g}_{3}$ and ${h}_{3}$ are of twist 4.
The leading-twist $\Phi_{\parallel, \perp} $ can be expanded as \cite{Chengt2010}
\begin{eqnarray}\label{t2}
\Phi_{(\parallel, \perp)}(u, \mu)= 6u (1-u)\sum_{\ell=1}^{\infty}\,a^{\ell}_{(\parallel, \perp)}(\mu)
C_{\ell}^{3/2}(\xi),
\end{eqnarray}
and twist-3 LCDAs  are related to twist-2
ones through the Wandzura-Wilczek relations:
\begin{eqnarray}\label{eq:WW}
g_v(u) &=& \int_{0}^u dv\, \frac{\Phi_\parallel(v)}{\bar v}+\int_{u}^1 dv\, \frac{\Phi_\parallel(v)}{v}\,,\nonumber\\
g_a(u) &=& 2\bar{u}\int_{0}^u dv\, \frac{\Phi_\parallel(v)}{\bar v}+ 2u\int_{u}^1 dv\, \frac{\Phi_\parallel(v)}{v}\,,\nonumber\\
h_t(u) &=& \frac{3}{2} (2u-1)\left(\int_{0}^u dv\, \frac{\Phi_\perp(v)}{\bar v} -\int_{u}^1 dv\, \frac{\Phi_\perp(v)}{v}\right)\,,
\nonumber\\
h_s(u) &=& 3 \left( \bar{u}\int_{0}^u dv\, \frac{\Phi_\perp(v)}{\bar v}+u\int_{u}^1 dv\, \frac{\Phi_\perp(v)}{v} \right)\,,
\end{eqnarray}
where $\mu$ is the normalization scale, $\bar{u}=1-u$ and $\xi=2u-1$.
Also, using the equation  of motion given in Ref. \cite{Ball99}, we can express the twist-4 DAs.

Two-parton  chiral-even light-cone distribution amplitudes of a
tensor meson are given by
\begin{eqnarray} \label{chialeven}
\langle T(p',\lambda)|\bar q_1(x)\gamma_\mu {q_2}(0)|0\rangle &=&
f_{T} m^2_T \int\limits_{0}^1 \! du\,e^{iup'x} \Bigg\{ p'_\mu
\frac{\varepsilon^{*\lambda}_{\alpha\beta}x^\alpha x^\beta}
{(p'x)^2} \,\Phi_\parallel^{T}(u) +
\frac{\varepsilon^{*\lambda}_{\mu\alpha} x^\alpha }{p'x}\,
g_v^{T}(u) - \frac{m_T^2}{2} x_\mu
\frac{\varepsilon^{*\lambda}_{\alpha\beta}x^\alpha x^\beta}{(p'x)^3}
\,g_3^{T}(u)\Bigg\}\,, \nonumber\\
\langle T(p',\lambda)|\bar q_1(x)\gamma_\mu\gamma_5 q_2 (0)|0\rangle
&=& f_{T} m_T^2 \int\limits_{0}^1 \! du\,e^{iup'x}
\epsilon_{\mu\nu\alpha\beta} {x^\nu p'^\alpha}
\varepsilon_{\lambda}^{*\beta\delta}x_\delta\,{g_a^{T}(u)\over
p'x}\,,
\end{eqnarray}
and the chiral-odd LCDA is
\begin{eqnarray}
\langle T(p',\lambda)|\bar q_1(x)\sigma_{\mu\nu}  q_2(0)|0\rangle
&=& -i f_{T}^\perp m_T \int\limits_{0}^1 \! du\,e^{iup'x}
\Bigg\{\left[\varepsilon^{*\lambda}_{\mu\alpha} x^\alpha p'_\nu -
\varepsilon^{*\lambda}_{\nu\alpha} x^\alpha p'_\mu\right]
\frac{1}{p'x} \Phi_\perp^{T}(u)
+(p'_\mu x_\nu - p'_\nu x_\mu) \nonumber\\
&& \times \frac{ m_T^2\varepsilon^{*\lambda}_{\alpha\beta}x^\alpha
x^\beta}{(p'x)^3} h_t^{T}(u) +\frac{m_T^2}{2}
\left[\varepsilon^{*\lambda}_{\mu\alpha} x^\alpha x_\nu -
\varepsilon^{*\lambda}_{\nu\alpha} x^\alpha x_\mu\right]
\frac{h_3^{T}(u)}{(p'x)^2}  \Bigg\} \,,
\end{eqnarray}
where $f_{T}$ is  scale independent, and $f_{T}^{\perp}$ is a scale-dependent decay constant
of the tensor meson $T$, as defined in Ref. \cite{Chengt2010}.

Now, two-parton distribution amplitudes should be inserted in Eq. (\ref{eqth1}),
and traces and integrals should be calculated. Finally,
 the same structures are equated both phenomenological and theoretical
sides of the correlation functions, and the Borel transform is performed
with respect to the variable $p^2$ as
\begin{eqnarray}\label{borel}
B_{p^2}(M^2)\frac{1}{\left(
p^{2}-m_{D}^{2}\right)^{n}}&=&\frac{(-1)^{n}}{\Gamma(n)}\frac{e^{-\frac{m_{D}^{2}}{M^{2}}}}{(M^{2})^{n}},
\end{eqnarray}
the sum rules are obtained for the form factors describing $D \to T$ decay.
 For instance, the form factor $A_{1}(q^{2})$ is obtain as
\begin{eqnarray}\label{eq3a1}
A_{1}(q^{2})&=&
-\frac{\alpha_{1}}{m_{T}+ m_{D}}\,
\Bigg\{\frac{f_{T}^{\perp}}{m_{T}}\int_{u_0}^{1}du\,\frac{\Phi_{\perp}^{(i)T}}{8 u^2}
\Bigg[-31 u+ 33\,(1+\frac{\delta_{1}(u)}{M^2})\Bigg] e^{-s(u)}
-f_{T}^{\perp}m_{T}\int_{u_0}^{1}du\,\frac{\bar{h}_{3}^{(ii)T}}{u^2 M^{2}} \Bigg[
\frac{\delta_{2}(u)}{M^{2}}-\frac{1}{2}
\Bigg]e^{-s(u)}\nonumber\\
&+&8\,f_{T}^{\perp}m_{T}\int_{u_0}^{1}du
\frac{h_{t}^{(iii)T}}{u^3 M^{2}} \Bigg[u-4+\frac{2\delta_{1}(u)}{M^{2}}
\Bigg]e^{-s(u)}+
f_{T} m_{c}\int_{u_0}^{1}du \frac{\left[-g_{v}^{(i) T}(u)+\frac{1}{2}g_{a}^{(i) T}(u)+8 \frac{m_{T}^{2}}{u M^{2}}\bar{g}_{3}^{(iii) T}
\right]}{u^2 M^{2}}e^{-s(u)}
\nonumber\\
&+&f_{T}^{\perp}m_{T}\int_{u_0}^{1}du\,\frac{h_{s}^{(i) T}}{u^2
M^{2}} e^{-s(u)}-
f_{T}^{\perp}m_{T}\int_{u_0}^{1}du\,\frac{h_{3}^{(ii) T}}{u^2 M^{2}}
\Bigg[5+\frac{m_{c}^2}{u\,M^{2}} \Bigg]e^{-s(u)} \Bigg\},
\end{eqnarray}
where
\begin{eqnarray*}
\begin{array}{ll}
u_{0}     =\frac{1}{2\, m_{T}^2} \left[\sqrt{(s_0-m_{T}^2-q^2)^2 +4
m_{T}^2 (m_c^2-q^2)} -\left(s_0-m_{T}^2-q^2\right)\right],         &                 \\
\alpha_{1}
=\frac{m_{T}^{2}~(m_{c}+m_{q})}{f_{D}\,m_{D}}\,e^{\frac{m_{D}^{2}}{M^{2}}},
                                                                    &
s(u)      =\frac{1}{u\, M^2}\left[m_c^2+u\,\bar{u}\,m_{T}^2-\bar{u}\, q^2\right],   \\
\delta_{1}(u) =\frac{m_c^2}{u}+m_{T}^{2}\,u+ q^{2} (u-2),          &
\delta_{2}(u) =\frac{m_c^2}{u}+ q^{2}\frac{\bar{u}^{2}}{u},                       \\
{f}^{(i)}(u)\equiv\int_0^u f(v) dv,                                &
{f}^{(ii)}(u)\equiv\int_0^u dv\int_0^v d\omega~ f(\omega),                        \\
{f}^{(iii)}(u)\equiv\int_0^u dv\int_0^v d\omega~ \int_0^\omega
d\tau~f(\tau).
\end{array}
\end{eqnarray*}

The explicit expressions for the other form factors are presented in
Appendix \ref{app:form factors}.

\section{Numerical analysis}\label{se.nu}
In this section, our numerical analysis of the sum rules for the
form factors and branching ratios is presented. In the calculation
of the form factors $V $ and $A_{i}(i=0,1,2)$, masses are taken in
$\rm {GeV}$ as $m_c=1.28\pm 0.03$, $m_{{D}^{0}}=1.86$,
$m_{{D}^{+}}=1.86$,  $m_{{D}^{+}_{s}}=1.96$ \cite{pdg}. For the $s$ and
$d$ quark at $\mu= 1 ~\rm{GeV}$, we take $m_{d}=(3.5-6) ~\rm{MeV}$
and $m_{d}=(104^{+26}_{-34})~\rm{MeV}$ \cite{Huang2004}. For
$D^{0}$, $D^{+}$, and $D_s$-meson decay constants,  the results of the
QCD sum rule as $f_{D^{0}}=f_{{D}^{+}}=(210.25 \pm
{11.60})~\mbox{MeV}$ and $f_{D_s}= (245.70\pm  7.46)~\mbox{MeV}$
\cite{Mutuk} are used.

For the tensor mesons,  the relevant parameters are presented in
Table \ref{Tenor}. All of the masses presented in Table \ref{Tenor}
are chosen from  Ref. \cite{pdg}, while the decay constants  $f_{T}
(f_{T}^{\perp})$ and the Gegenbauer moments $a^{1}_{(\parallel,
\perp)}$ are taken from Ref. \cite{Cheng2010}.

\begin{table}[th]
\caption{Masses, decay constants and Gegenbauer moments for tensor  mesons. The
decay constants and Gegenbauer moments  are given at the scale $\mu=1 ~\rm {GeV}$.} \label{Tenor}
\begin{ruledtabular}
\begin{tabular}{ccccc}
$T$ &$a_2$&$K_2^{*}$&$f_{2}$&$f_{2}'$\\
\hline
Mass\,(GeV)&$1.32$&$1.42$&$1.27$&$1.52$\\
$f_{T}$ (MeV)&$107\pm 6$ &$118\pm 5$&$102\pm 6$&$126\pm4$\\
$f_{T}^{\perp}$ (MeV)&$105\pm21$&$77\pm 14$&$117\pm 25$&$65\pm12$\\
$a^{1}_{(\parallel, \perp)}$&$5/3$&$5/3$&$5/3$&$5/3$\\
\end{tabular}
\end{ruledtabular}
\end{table}

\subsection{Analysis of the form factors}
\begin{figure}[!h]
\includegraphics[width=6cm,height=6cm]{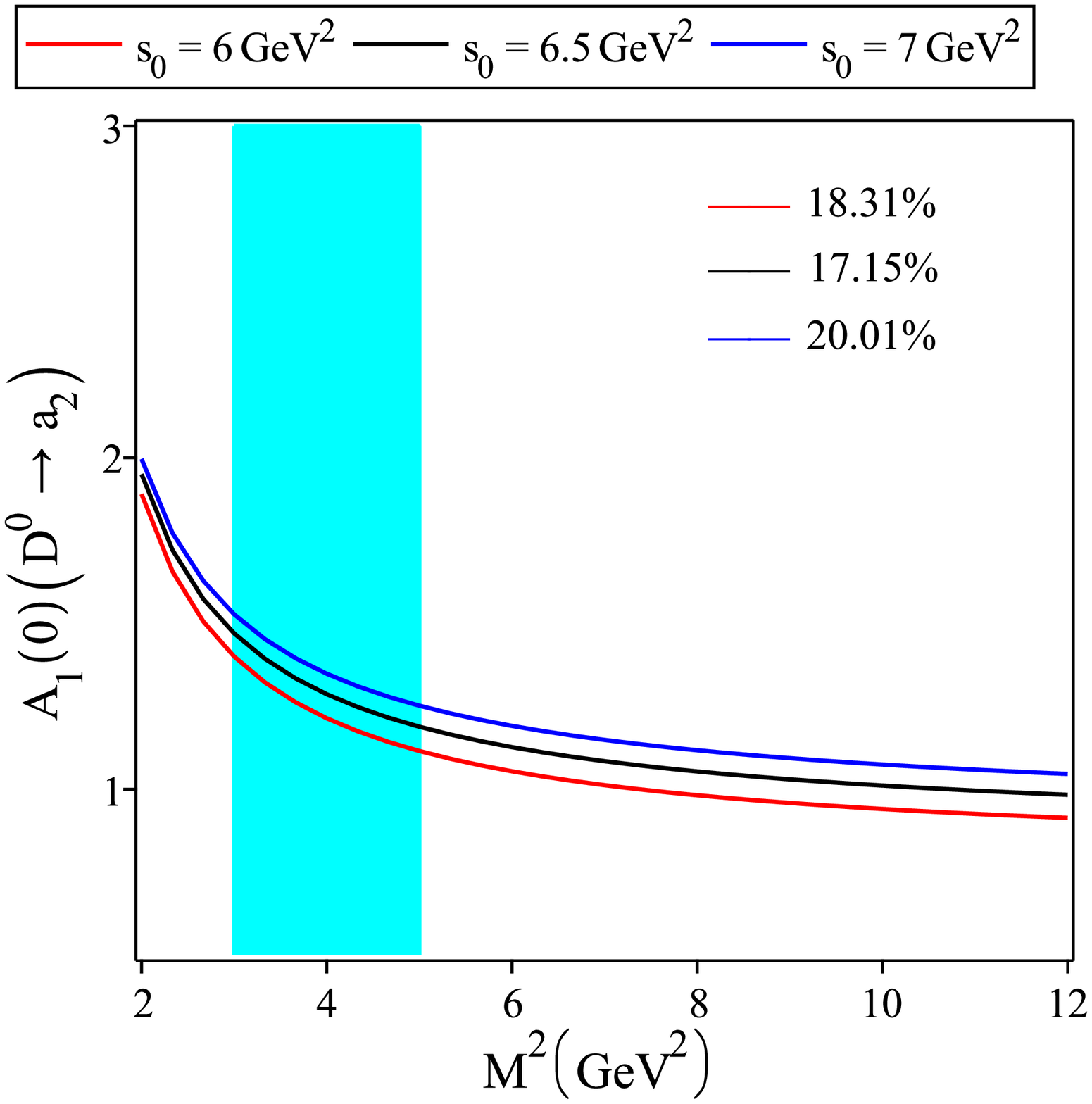}
\includegraphics[width=6cm,height=6cm]{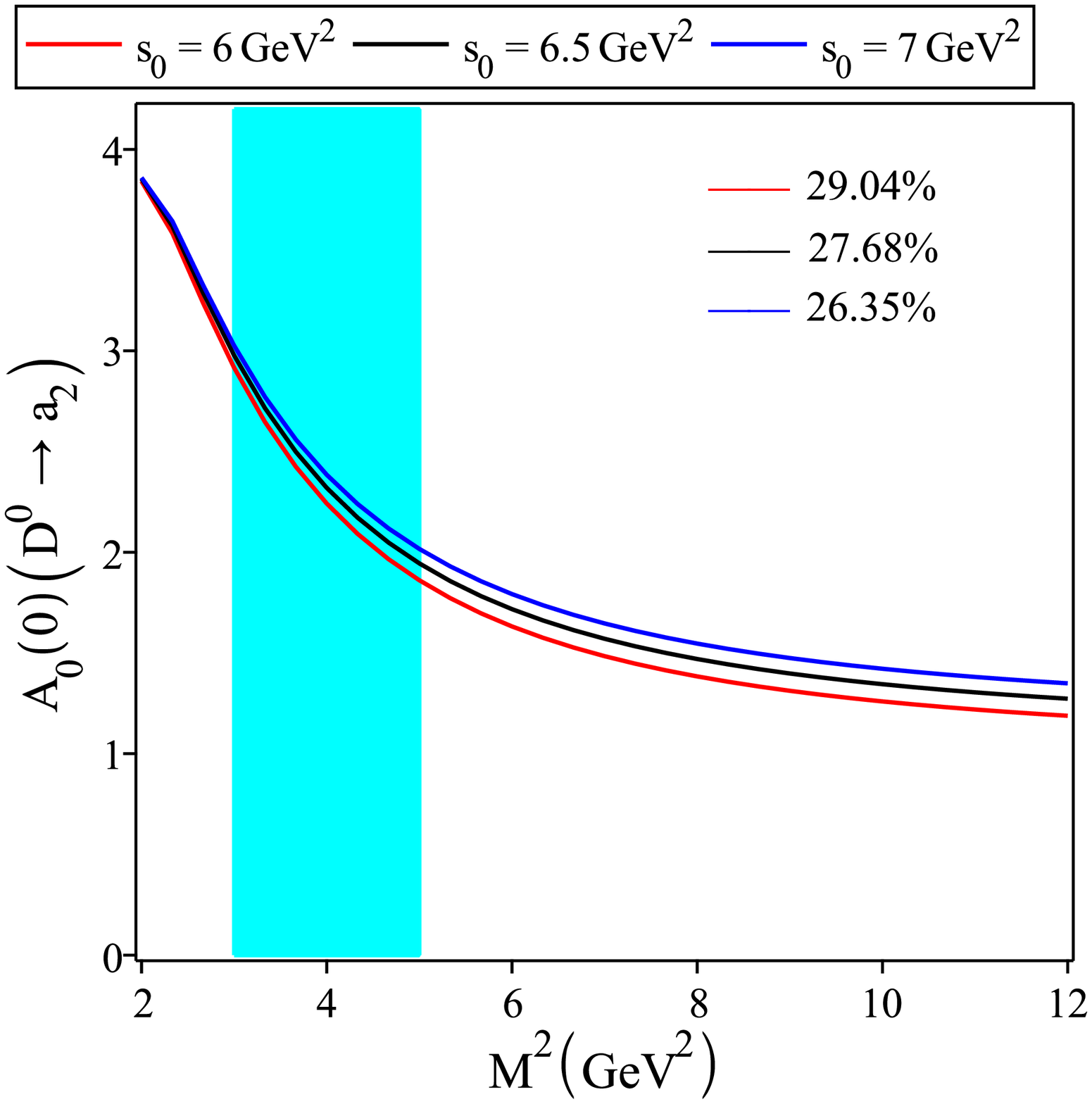}
\includegraphics[width=6cm,height=6cm]{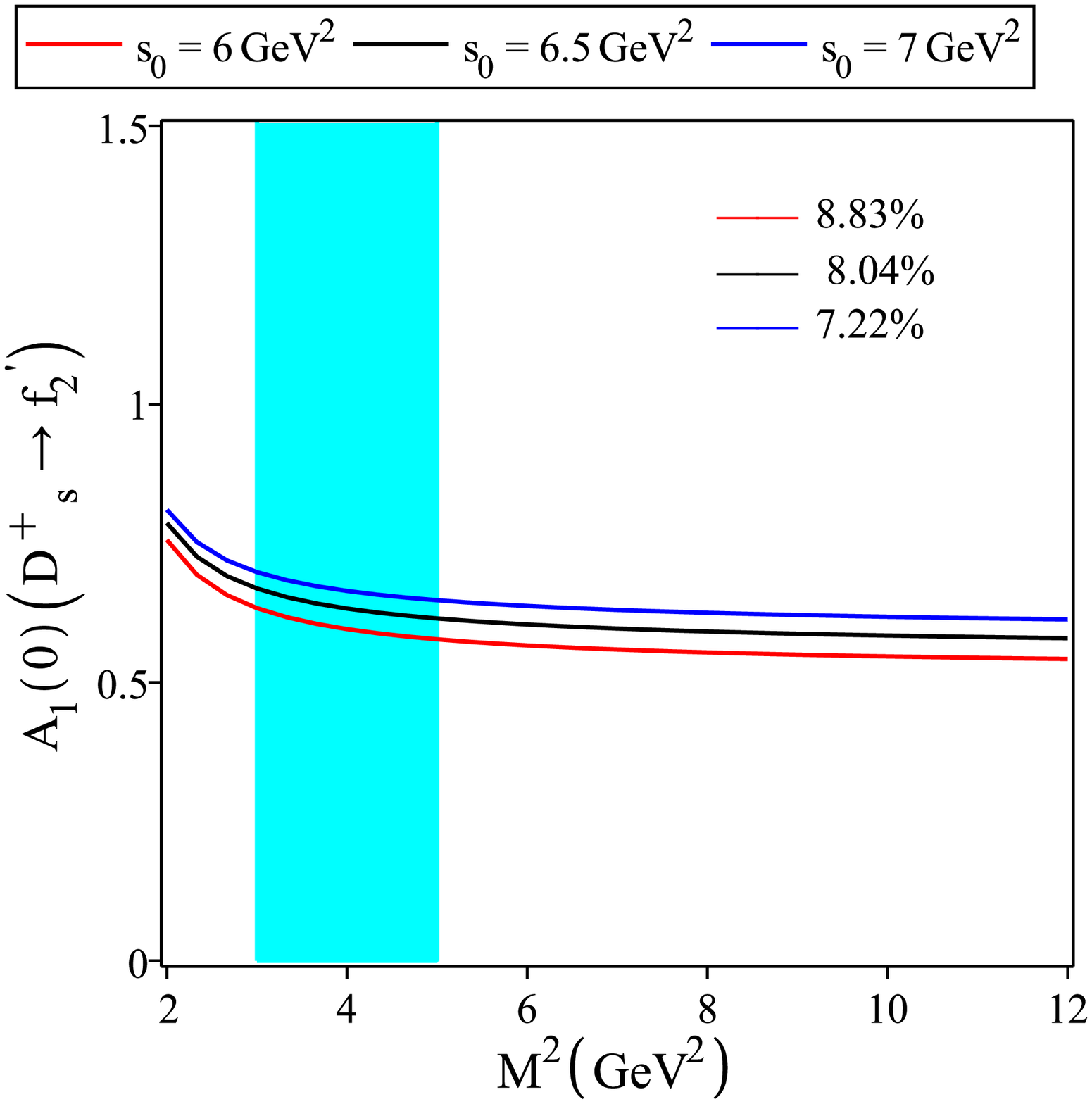}
\includegraphics[width=6cm,height=6cm]{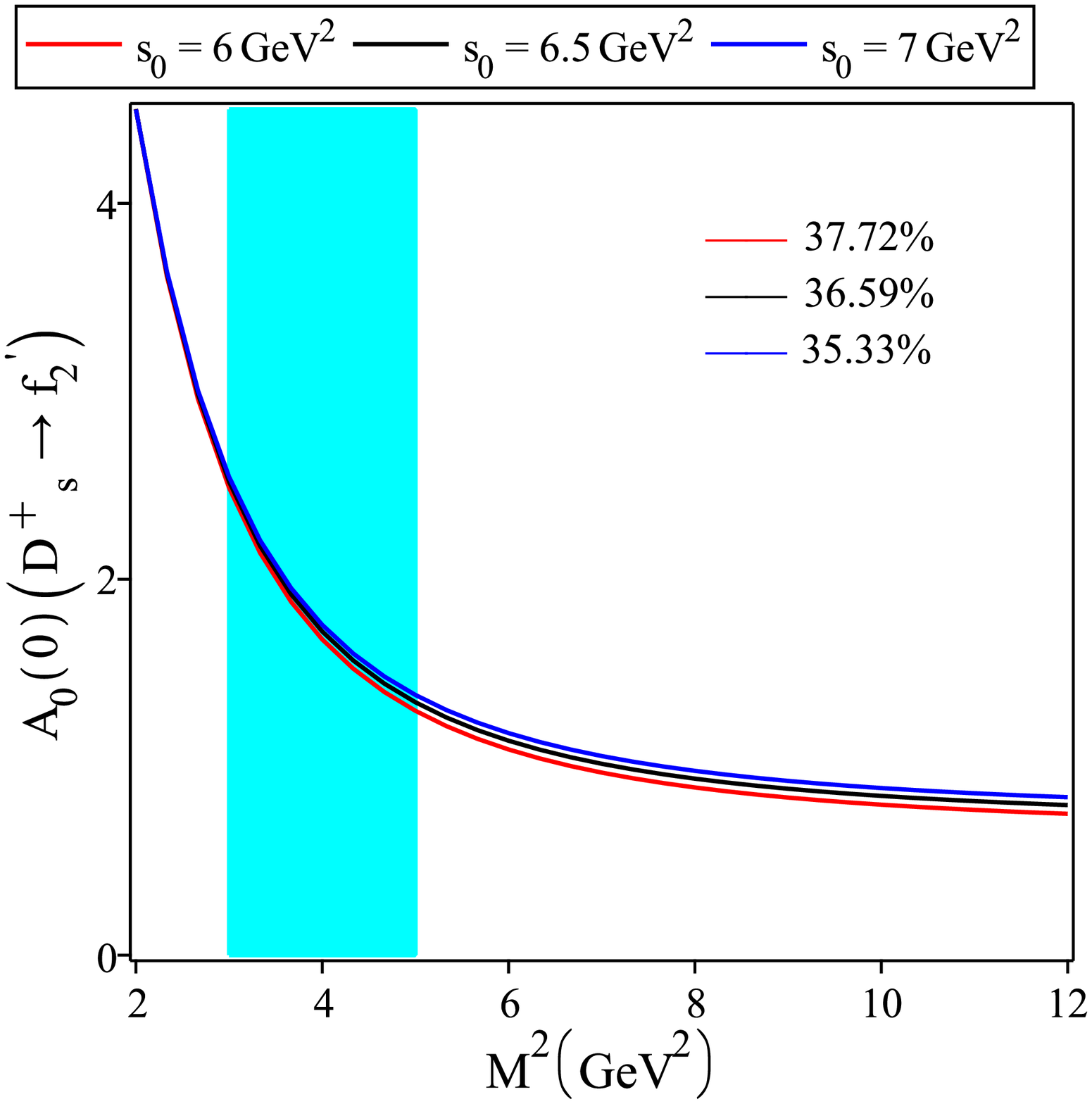}
\caption{The form factors $A_{1}(0)$ and $A_{0}(0)$ for the
$D^{0}(D_{s}^{+})\to a_{2}(f_{2}')$ transition on $M^{2}$ at three
values of the threshold $s_{_0}$. The the relative change in the
value of the form factors at the shaded interval of the Borel
parameter is displaced in every plot. } \label{F1}
\end{figure}

In this subsection,  our numerical analysis of the form factors is
presented. The sum rules for the form factors  contain two
parameters: namely, Borel mass squares $M^{2}$ and continuum
thresholds $s_{0}$. Our results should be independent of these
parameters since $M^{2}$ and $s_{0}$ are not physical quantities. In
this paper the value of continuum threshold is used as $s_{0}\in [6,
8] ~ \mbox{GeV}^2$ \cite{Khosravi}. To carry out numerical
calculations,  a region of $M^{2}$ must be obtained and  the
suitable region has two conditions. First, the nonperturbative terms
must remain subdominant by the lower bound of $M^{2}$;  and second,
the higher bound must decrease the contributions of the higher
states and continuum. In Fig. \ref{F1},  the $M^{2}$ dependence of
the form factors $A_{1}(q^2=0)$ and $A_{0}(q^2=0)$ is presented for
$D^{0}(D_{s}^{+})\to a_{2}(f_{2}')$ transition, at three different
values of the threshold $s_{0}=6~ \rm{GeV^2} $, $s_{0}=6.5~
\rm{GeV^2} $ and $s_{0}=7~ \rm{GeV^2} $, with red, black,  and blue
lines, respectively. In this figure, the relative change in the value
of the form factors at $q^2=0$ is also displaced at the shaded interval
of the Borel parameter. Our numerical analysis  reveals that for $3~
\rm{GeV^2} \leq M^{2} \leq 5~ \rm{GeV^2}$  all of the form factors
show good stability.

\begin{figure}[!h]
\includegraphics[width=7.5cm,height=7cm]{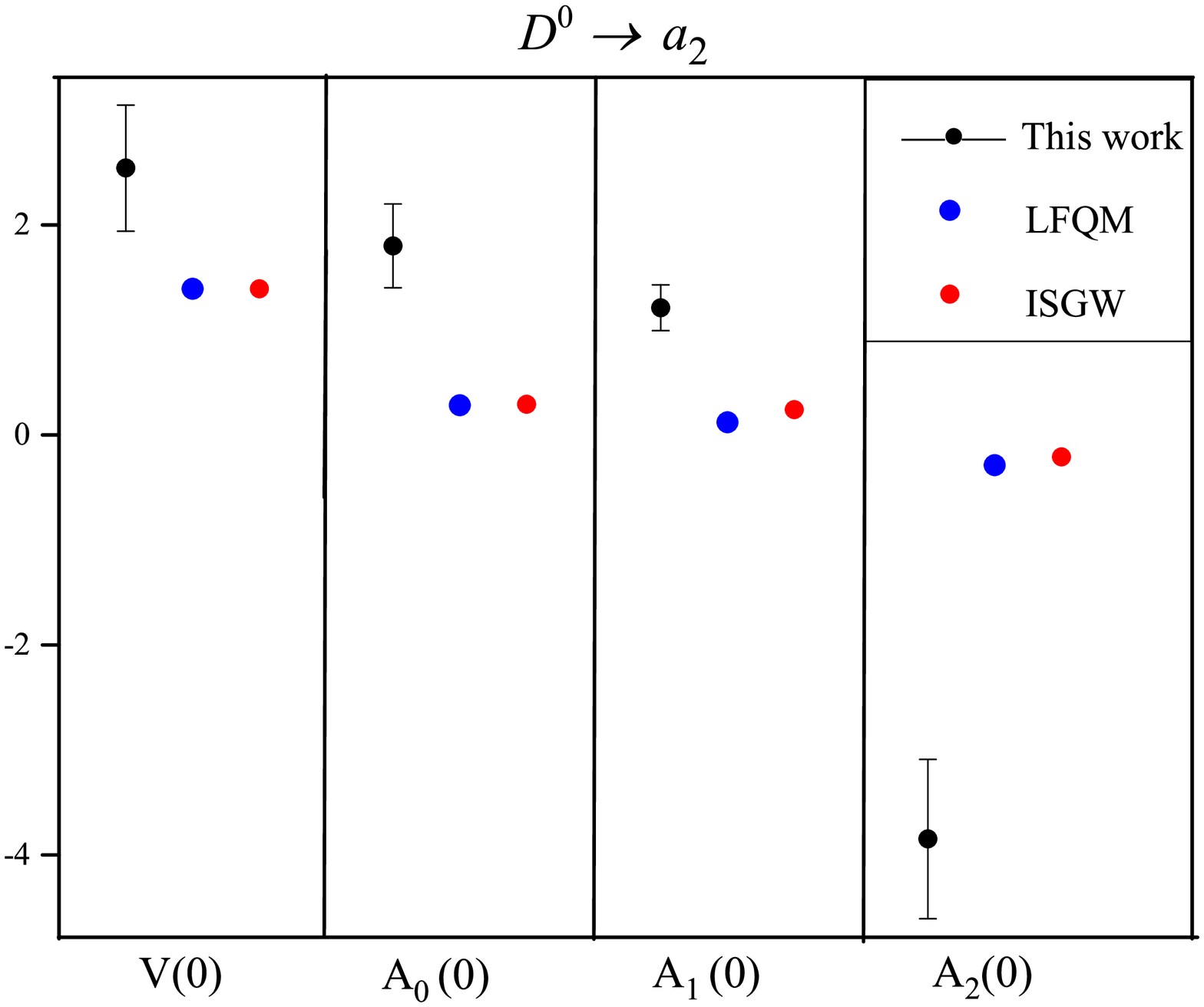}
\includegraphics[width=7.5cm,height=7cm]{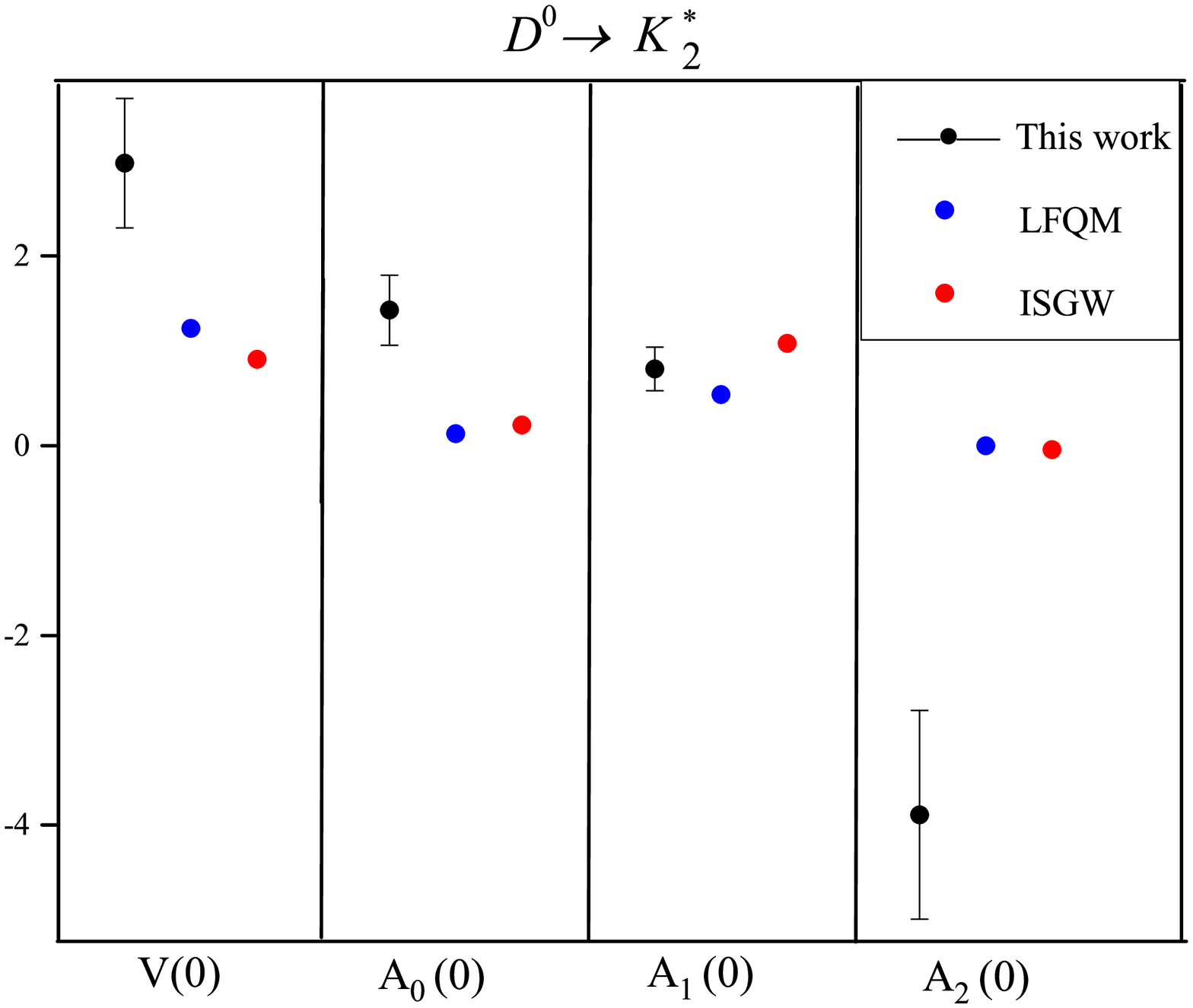}
\includegraphics[width=7.5cm,height=7cm]{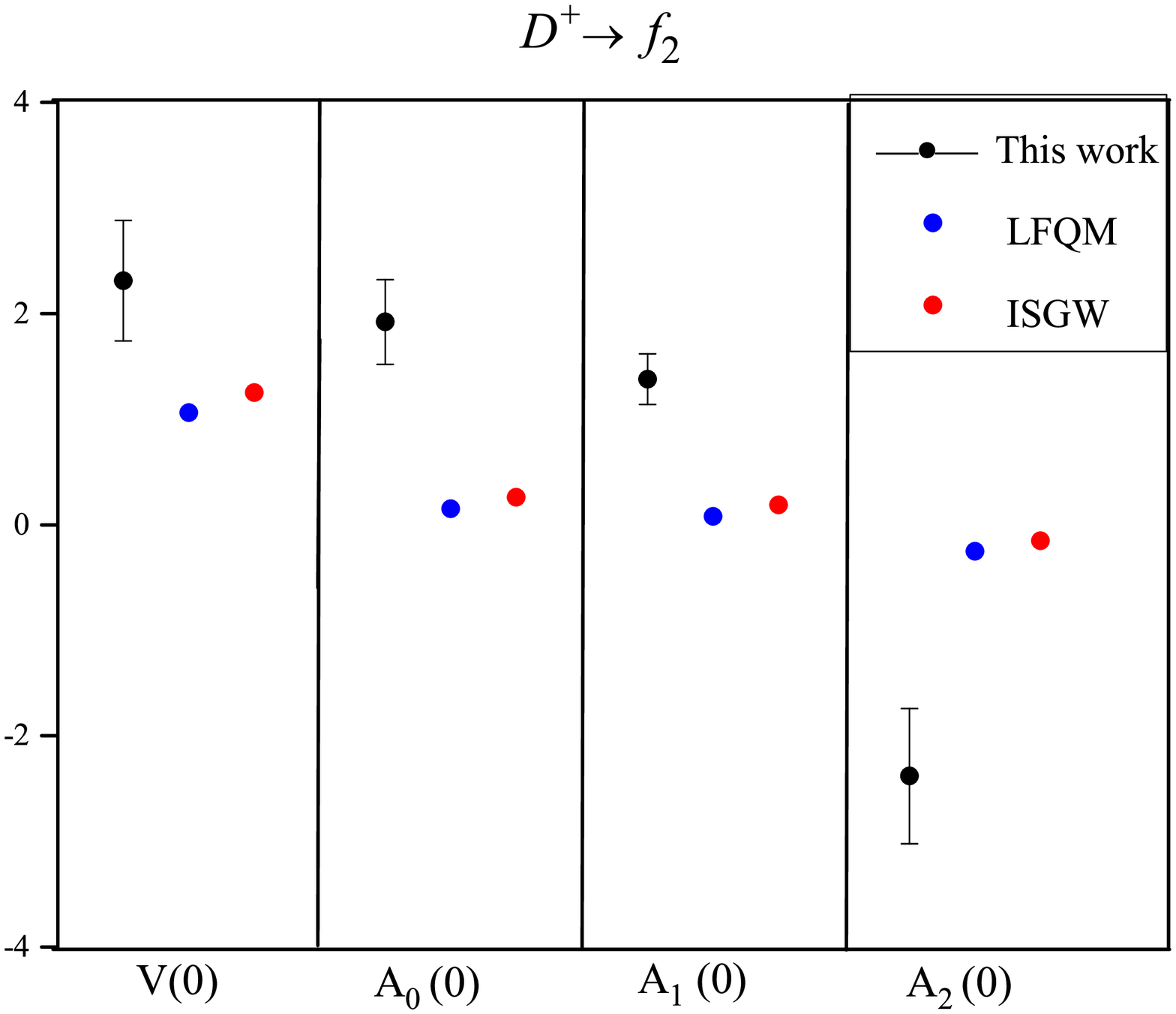}
\includegraphics[width=7.5cm,height=7cm]{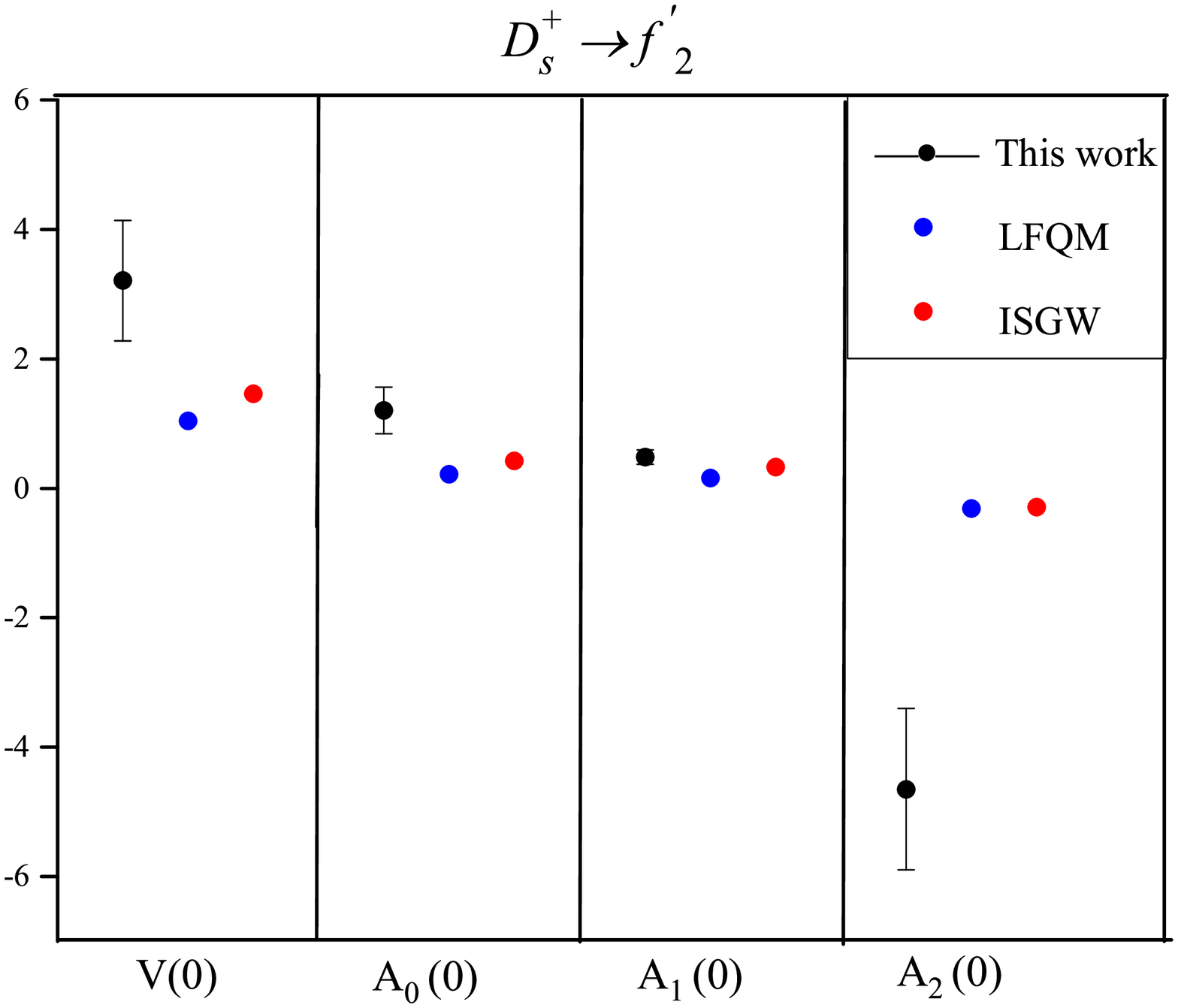}
\caption{The values of the ${V}(0), {A}_{i}(0)$ in comparison with
the predictions of the other approaches, such as  LFQM and ISGW.  }
\label{compff}
\end{figure}

Now, the $q^2$ dependency of the form factors can be carried out.
First, the values of the form factors at $q^2=0$ are estimated. In
Fig. \ref{compff}, our results for ${V}, {A}_{i}(i=0, 1, 2)$ of $D
\to T \ell \bar{\nu}_{\ell}$ decays in $q^2=0$ are presented.
Moreover, this table contains the predictions of the covariant
light-front model (LFQM) and improved version ISGW quark model
 approaches \cite{Chengb,Scora1995}. The results of the other
approaches are rescaled according to Eq. (\ref{eqffp}). The errors
in Fig. \ref{compff} are estimated by the variation of the Borel
parameter $M^{2}$ and the decay constants $f_{T}$ and
$f_{T}^{\perp}$.

\begin{figure}[!h]
\includegraphics[width=6cm,height=6cm]{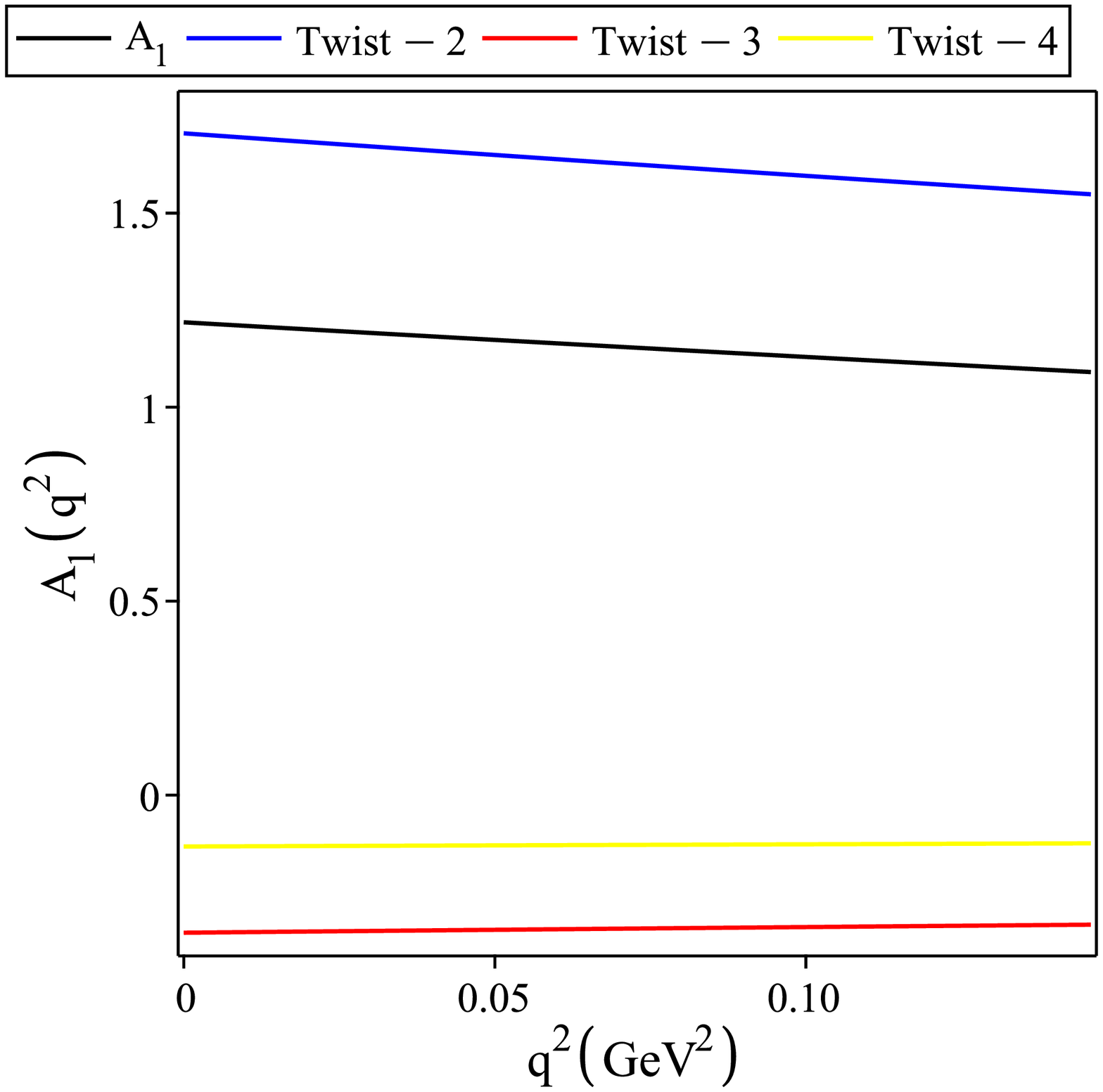}
\includegraphics[width=6cm,height=6cm]{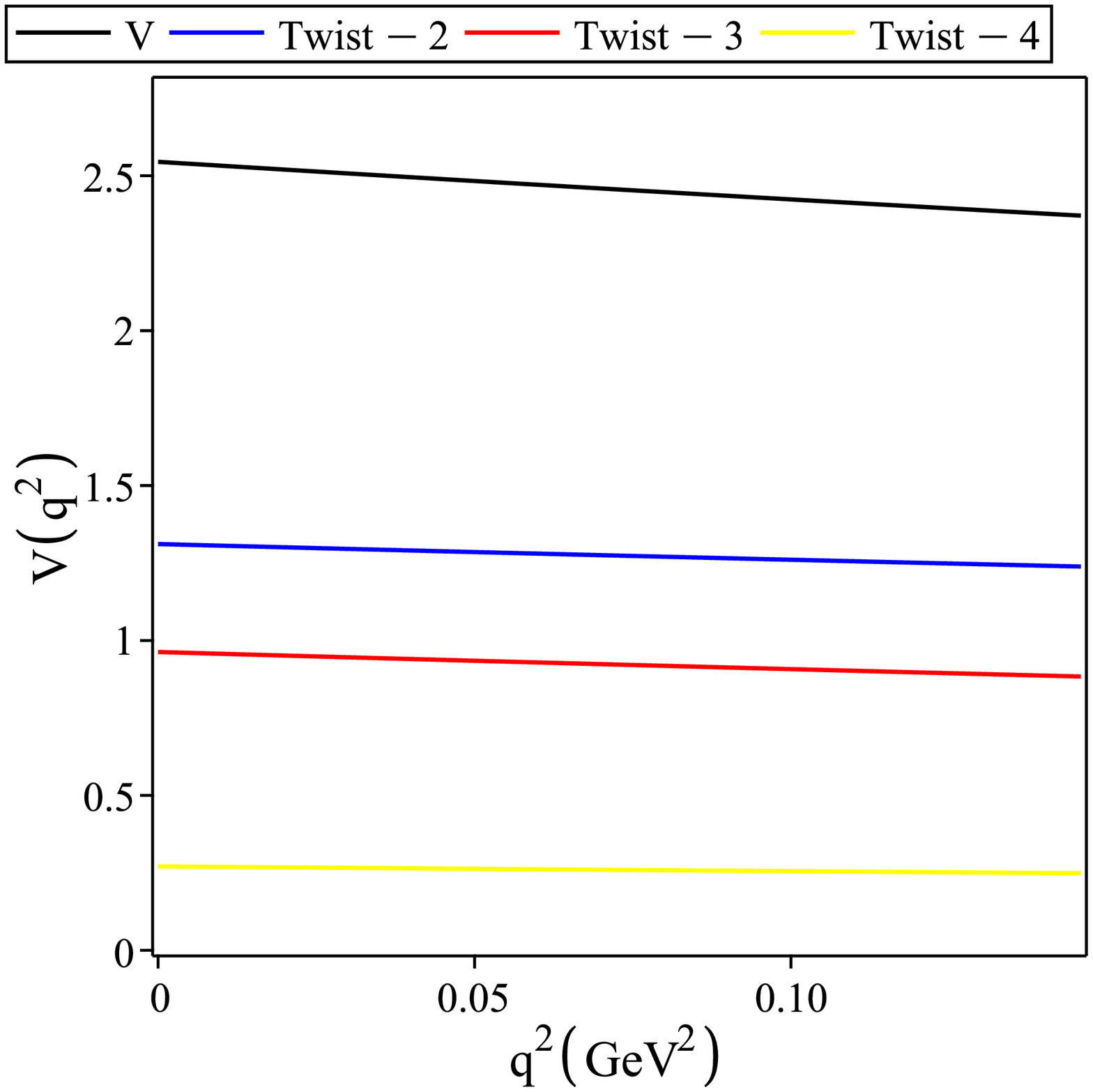}
\caption{ $A_{1}$ and $V$ form factor of $D^{0}\to a_{2}$ transition
on $q^2$. The contributions of twist-2, 3 and twist-4 functions in
these form factors are displaced with blue, red, and yellow lines,
respectively. } \label{F13}
\end{figure}

Figure \ref{F13} depicts the twist-2,3 and twist-4 contributions in
the form factor formula $A_{1}(q^2)$ and $V({q^2})$ for $D^{0}\to
a_{2}$ decay. Similarity, as shown in this figure, for all of the
form factors the most contribution is related to the twist-2 DAs,
while the twist-4 DAs have the least contribution.

\begin{figure}[!h]
\includegraphics[width=6cm,height=6cm]{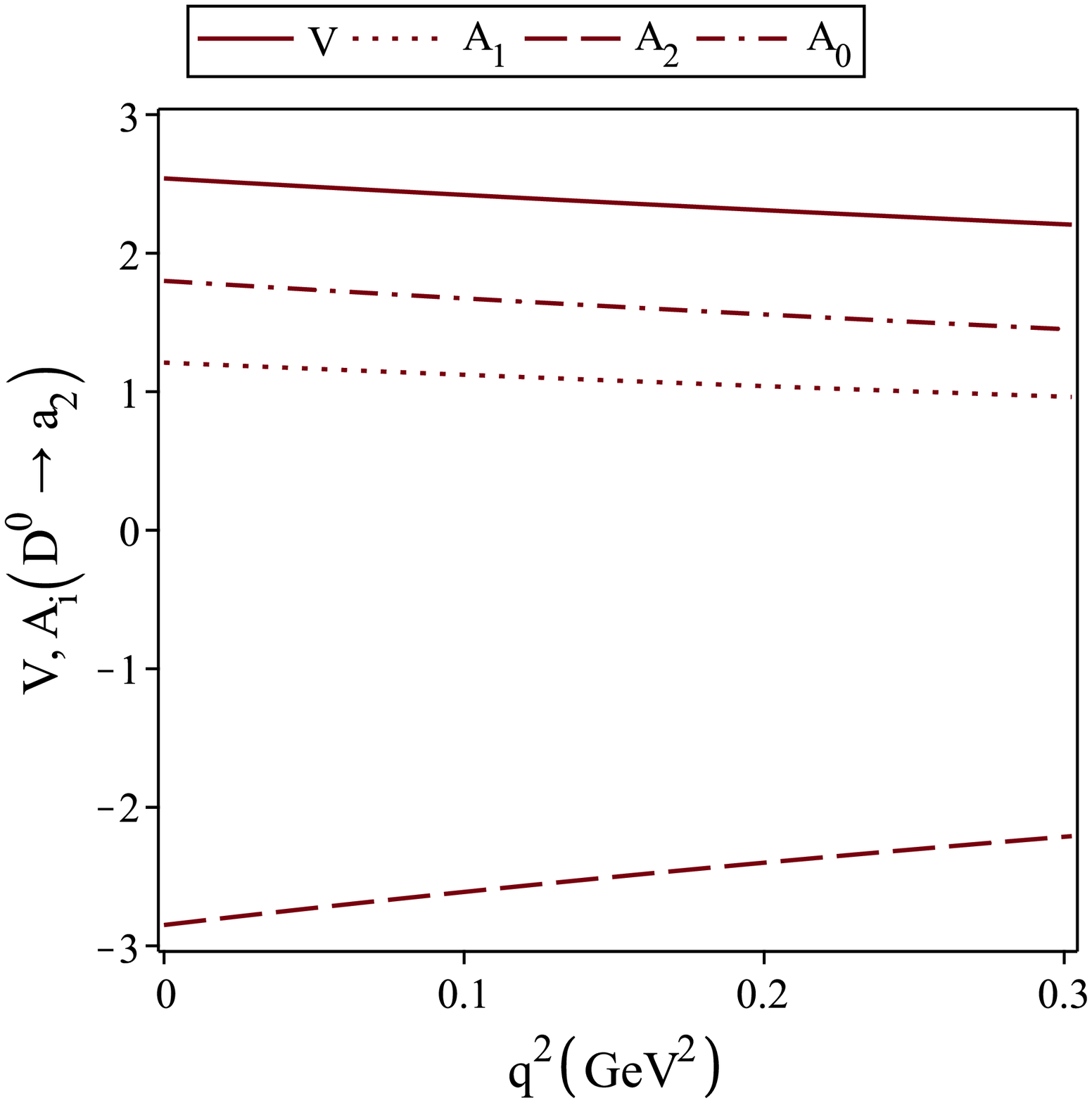}
\includegraphics[width=6cm,height=6cm]{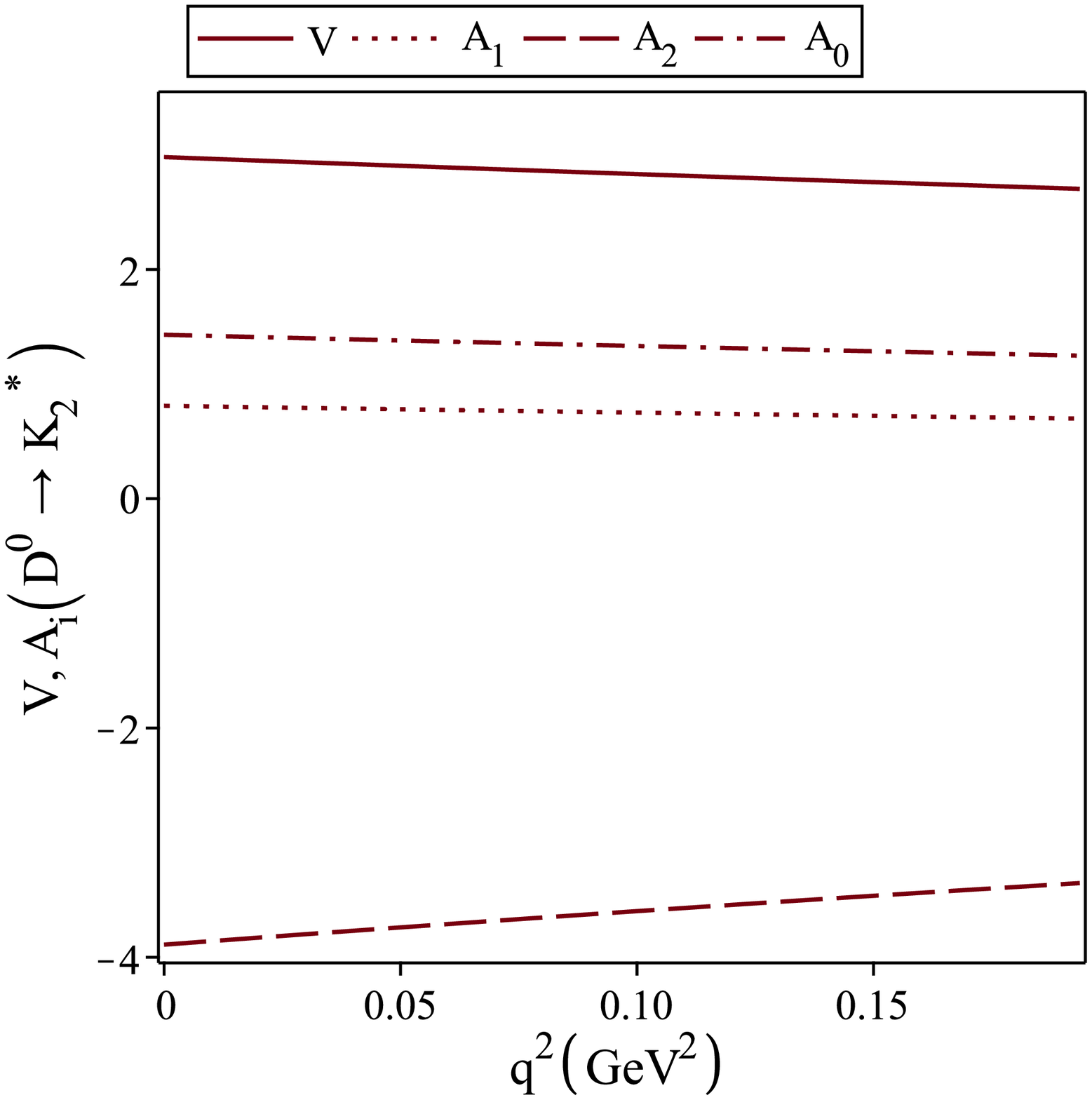}
\includegraphics[width=6cm,height=6cm]{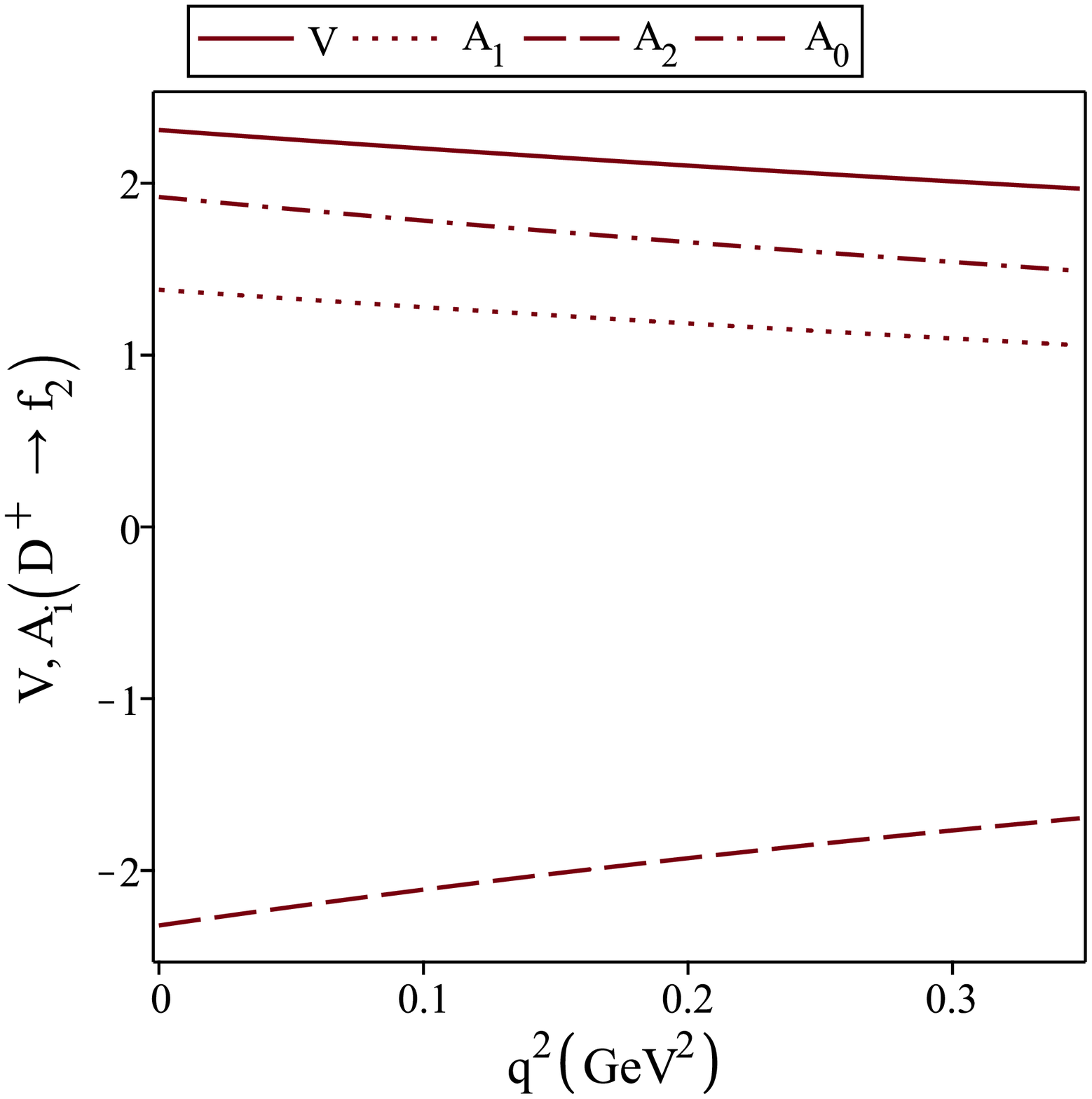}
\includegraphics[width=6cm,height=6cm]{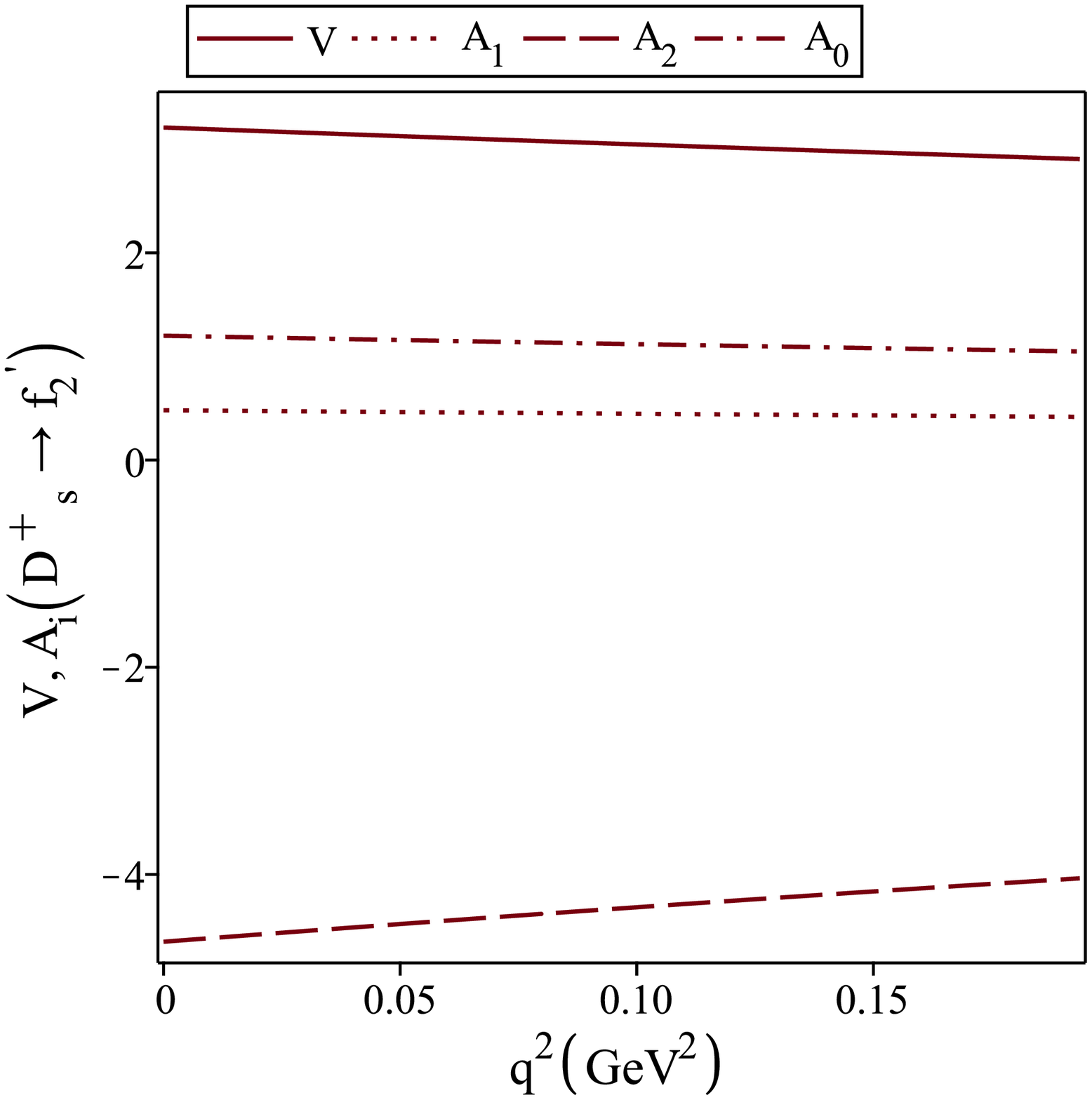}
\caption{The form factors $V, A_{i}$   on $q^2$ for $D\to T$ decay. }
\label{ffdtot}
\end{figure}

To extend the present result to the whole physical region,
$m_\ell^2 \le q^2 \le (m_D-m_{T})^2$, we use the
parametrization of the form factors with respect to $q^2$ as
\begin{eqnarray}\label{fitt}
F(q^{2})=\frac{F(0)}{1+\alpha\,q^2/m_{D}^{2}+\beta\,q^4/m_{D}^{4}},
\end{eqnarray}
where  $F({0})$ denotes the value of the form factor at $q^2=0$. In
addition, $\alpha$ and $\beta$ are  the corresponding fitting
coefficients listed in Table \ref{fittp} for different form factors.
\begin{table}[th]
\caption{The values of the parameter $F(0)$, $\alpha$, and
 $\beta$ for each form factor.} \label{fittp}
\begin{ruledtabular}
\begin{tabular}{cccc|cccc}
$\mbox{Form factor}$& ${F(0)}$&$\alpha$&$\beta$ &$\mbox{Form factor}$& ${F(0)}$&$\alpha$&$\beta$\\
\hline
$V^{D^0\to a_{2}}$& ${2.54}$&$1.71$&$0.46$ &$V^{D^0\to K^{*}_{2}}$& ${2.98}$&$1.80$&$0.58$\\
$A_{1}^{D^0\to a_{2}}$& ${1.21}$&$2.61$&$4.22$ &$A_{1}^{D^0\to K^{*}_{2}}$& ${0.81}$&$2.60$&$4.43$\\
$A_{2}^{D^0\to a_{2}}$& ${-2.85}$&$3.12$&$2.75$ &$A_{2}^{D^0\to K^{*}_{2}}$& ${-3.89}$&$2.77$&$1.81$\\
$A_{0}^{D^0\to a_{2}}$& ${1.80}$&$2.56$&$2.75$ &$A_{0}^{D^0\to K^{*}_{2}}$& ${2.98}$&$1.80$&$0.58$\\
$V^{D^{+}_{s}\to f_{2}'}$& ${3.21}$&$2.04$&$0.81$ &$V^{D^{+}\to f_{2}}$& ${2.31}$&$1.68$&$0.42$\\
$A_{1}^{D^{+}_{s}\to f_{2}'}$& ${0.48}$&$2.86$&$6.54$ &$A_{1}^{D^{+}\to f_{2}}$& ${1.38}$&$2.62$&$4.14$\\
$A_{2}^{D^{+}_{s}\to f_{2}'}$& ${-4.65}$&$2.91$&$1.94$ &$A_{2}^{D^{+}\to f_{2}}$& ${-2.32}$&$3.32$&$3.35$\\
$A_{0}^{D^{+}_{s}\to f_{2}'}$& ${1.20}$&$2.73$&$2.83$ &$A_{0}^{D^{+}\to f_{2}}$& ${1.92}$&$2.58$&$2.85$\\
\end{tabular}
\end{ruledtabular}
\end{table}
The dependence of the fitted form factors  $V, A_{i} (i=0, 1, 2)$ on
$q^2$ for $D\to T$ transitions is shown in Fig. \ref{ffdtot}.

\subsection{Differential branching ratio for the semileptonic decays}
Now, we would like to evaluate the branching ratio values for the $D\to T \ell \bar{\nu}_{\ell}$ decays.
The expressions of the differential decay width
are given as
\begin{eqnarray}\label{eq221}
\frac{d\Gamma_{L}(D\to T \ell \nu)}{dq^2}&=&
\sigma\, \upsilon^{2}\,
\left[\frac{1}{9}(2q^2+m_{\ell}^{2})h_{0}^{2}(q^{2})+\frac{1}{3}
\lambda m_{\ell}^{2} A_{0}^{2}(q^{2})\right],\nonumber\\
\frac{d\Gamma_{\pm}(D\to T \ell \nu)}{dq^2}&=&
\sigma\, \upsilon^{2}\,
\frac{(2 q^2 + m_{\ell}^{2})}{12}\Bigg[|(m_{D}+m_{T})A_{1}(q^{2})
\mp\frac{\sqrt{\lambda}}{m_{D}+m_{T}}V(q^2)|^{2}\Bigg],
\end{eqnarray}
where  $m_{\ell}$ represents the mass of the charged lepton and $L,
\pm $ denotes the helicities of the tensor mesons. The other
parameters are defined as
\begin{eqnarray}\label{eq222}
\sigma&=&\frac{G_{F}^{2}\,|V_{q'c}|^{2}\sqrt{\lambda^{3}(m_{D}^{2},m_{T}^{2},q^2)}}{256~
m_{D}^{5}~\,m_{T}^2~\pi^{3}q^{2}},~~~~~\upsilon=\left(1-\frac{m_{\ell}^{2}}{q^{2}}\right),\nonumber\\
h_{0}(q^{2})&=&\frac{1}{2 m_{T}}\left[(m_{D}^2 - m_{T}^2 -
q^{2})(m_{D}+m_{T})A_{1}(q^{2})-\frac{\lambda}{m_{D}+m_{T}}A_{2}(q^{2})\right].
\end{eqnarray}

\begin{figure}[!h]
\includegraphics[width=6cm,height=6cm]{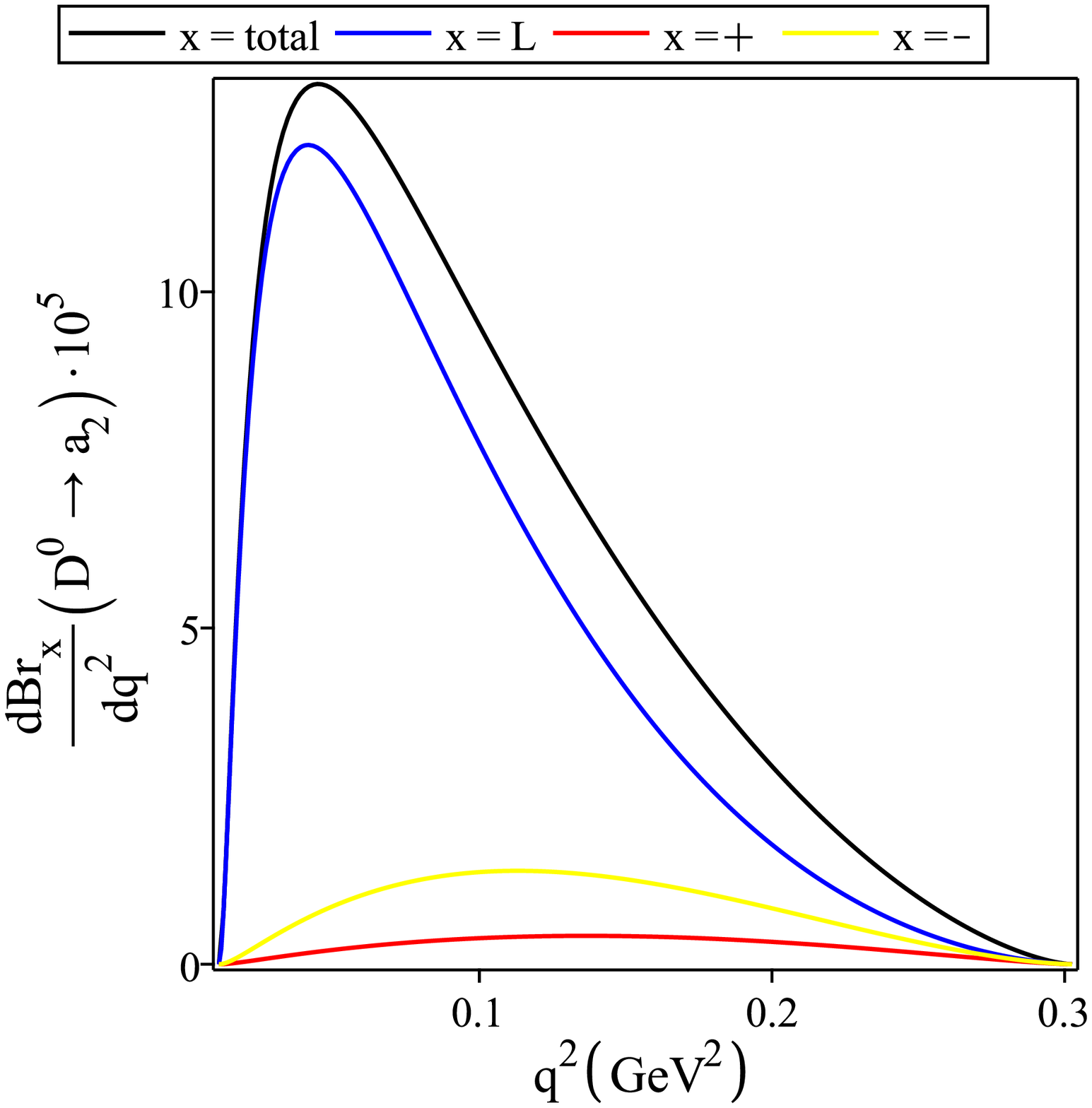}
\includegraphics[width=6cm,height=6cm]{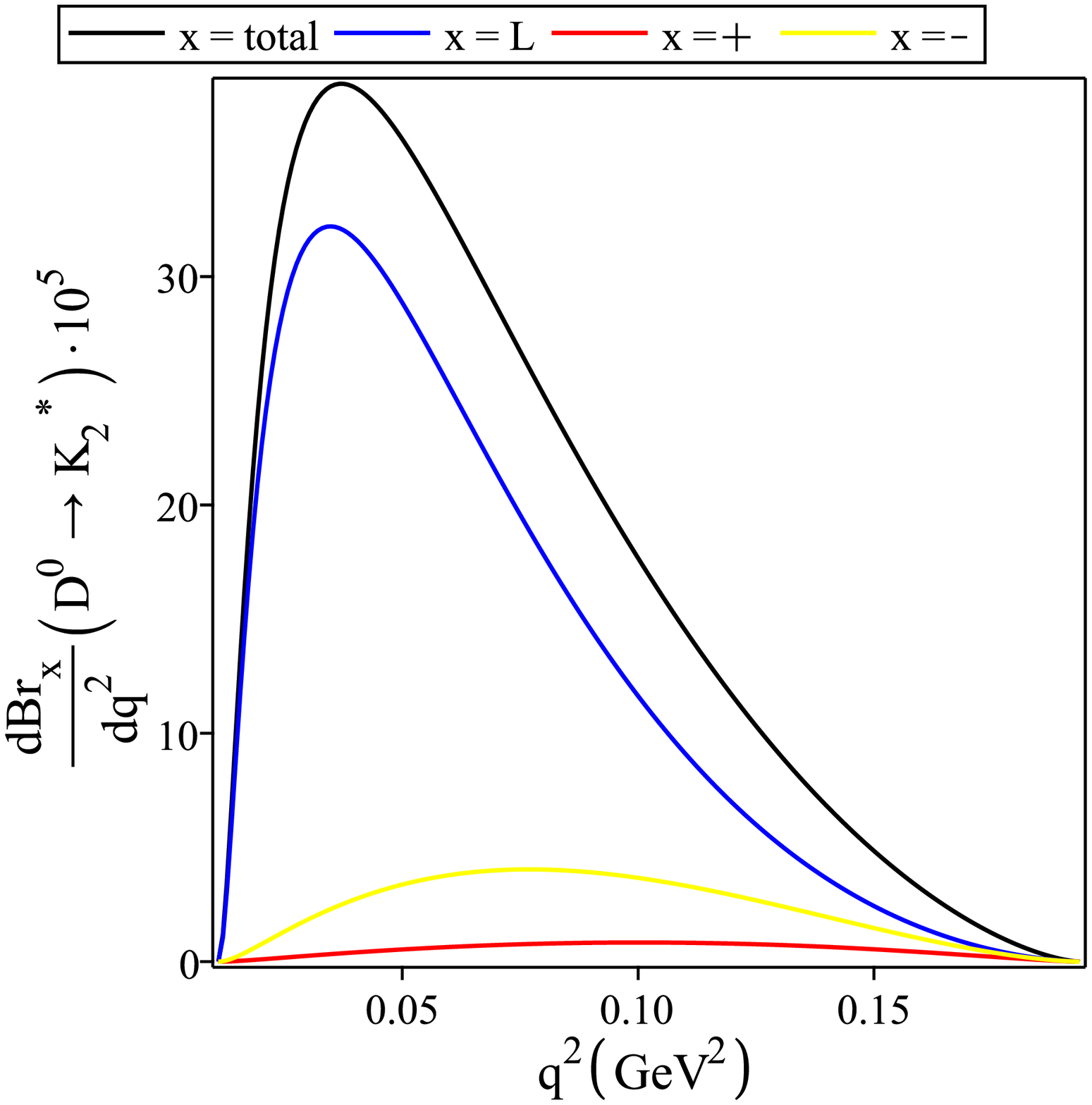}
\includegraphics[width=6cm,height=6cm]{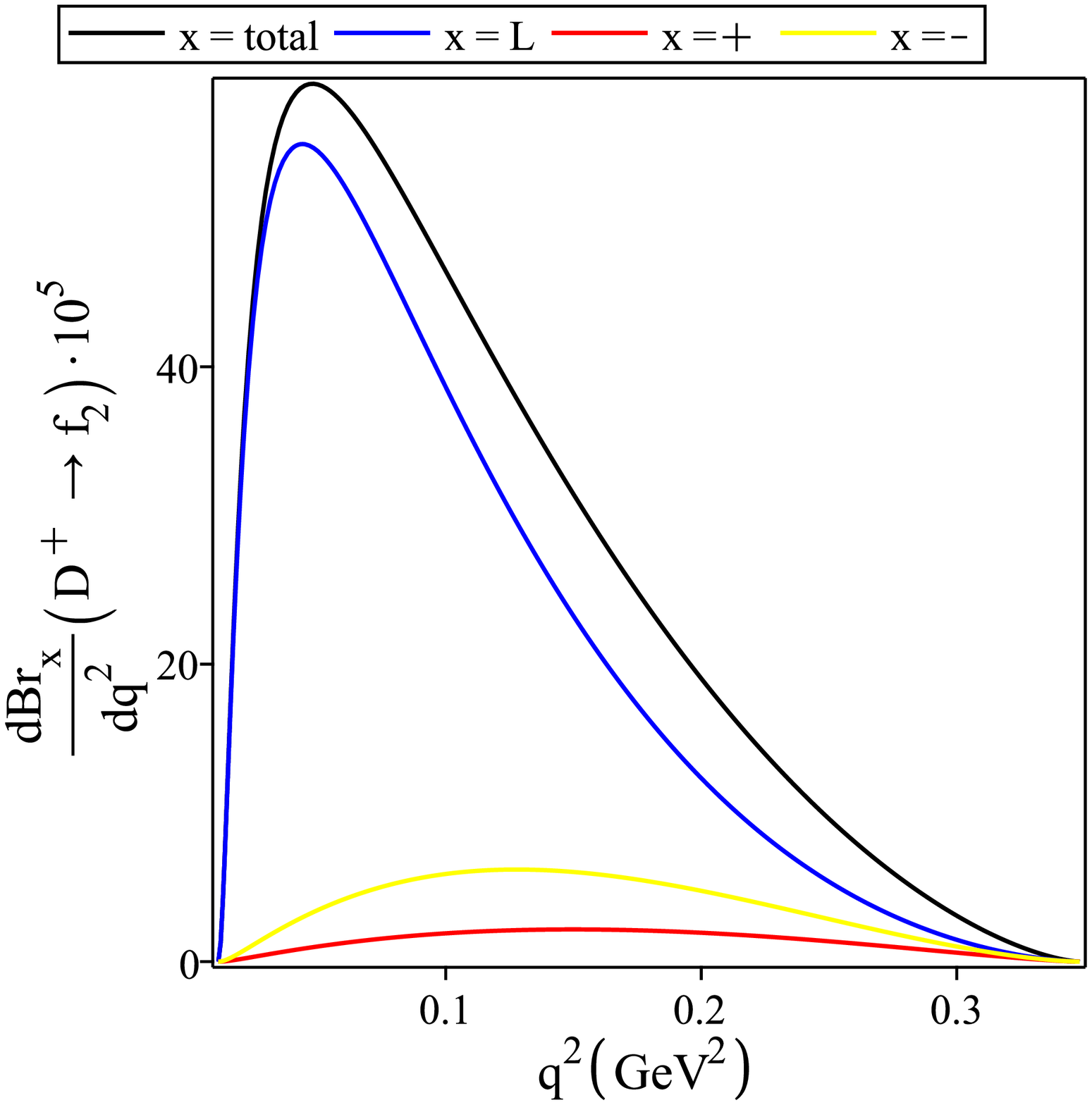}
\includegraphics[width=6cm,height=6cm]{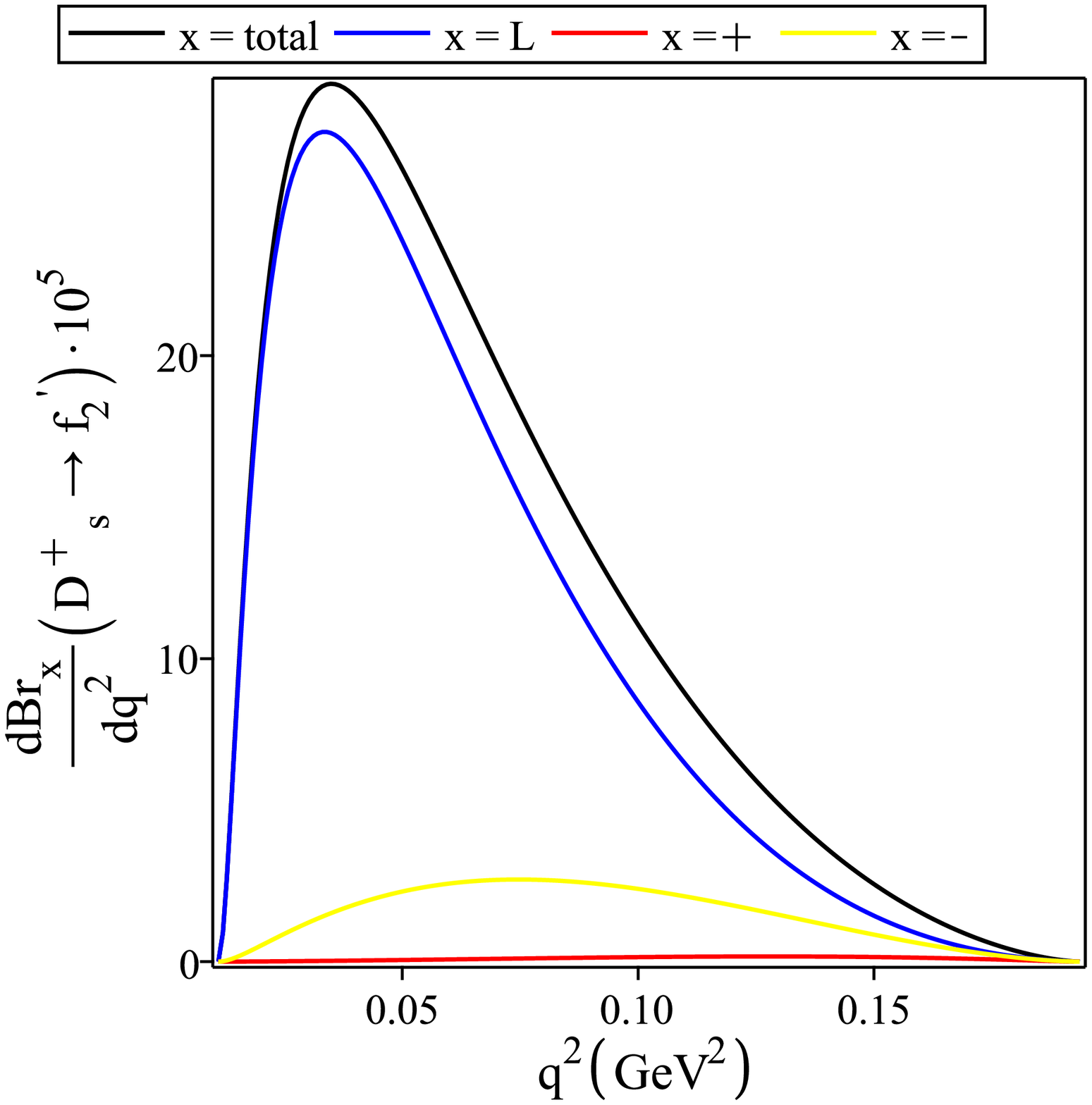}
\caption{Differential branching ratios  of the $D\to T\,
\mu\,\bar{\nu}_{\mu}$ as functions of $q^2$. } \label{F2}
\end{figure}

The differential branching ratios of the $D\to T\,
\mu\,\bar{\nu}_{\mu}$ decays are plotted on $q^2$ in Fig. \ref{F2},
in which take $|V_{cd}|=(0.22\pm 0.00)$, $|V_{cs}|=(0.98\pm 0.01)$
and $m_{\mu}=105.65 ~ \rm{MeV}$ \cite{pdg}.
In this figure, the black, blue, red, and yellow lines show
$dBr_{total}/dq^2$, $dBr_{L}/dq^2$, $dBr_{+}/dq^2$, and
$dBr_{-}/dq^2$, respectively. Integrating Eq. (\ref{eq221}) over
$q^2$ in the whole physical region and using the total mean
lifetime $\tau_{D^0}=0.41 $, $\tau_{D^+}=1.04$ and,
$\tau_{D^{+}_s}=0.50 $ ps \cite{pdg}, the branching ratio values of
these decays are  obtained as presented in Table \ref{Br}.
\begin{table}[th]
\caption{Branching ratios of $D\to T\, \mu\,\bar{\nu}_{\mu}$
obtained in this work. $\rm{Br}_{L}$, $\rm{Br}_{+}$, and
$\rm{Br}_{-}$ stand for the portion of the rate with a longitudinal
polarization, positive helicity,  and negative helicity of the $T$-meson,
respectively. The error comes from the variation of form factors. }
\label{Br}
\begin{ruledtabular}
\begin{tabular}{ccccc}
Decay&$[\rm{Br}_{L}$&$\rm{Br}_{+}$&$\rm{Br}_{-}$&$\rm{Br}_{total}]\times 10^{5}$ \\
\hline
$D^0\to a_{2}\,\mu\,\bar{\nu}_{\mu}$&$1.23 \pm 0.44$&$0.06\pm 0.01$&$0.21 \pm 0.06$&$1.50 \pm 0.51$ \\
$D^0\to K_{2}^{*}\,\mu\,\bar{\nu}_{\mu}$&$2.35 \pm 0.83$&$0.09 \pm 0.03$&$0.41 \pm 0.15 $&$2.85 \pm 1.01$ \\
$D^{+}_{s}\to f_{2}'\,\mu\,\bar{\nu}_{\mu}$&$1.52 \pm 0.73$&$0.02 \pm 0.01$&$0.22\pm 0.07$&$1.76 \pm 0.81$ \\
$D^{+}\to f_{2}\,\mu\,\bar{\nu}_{\mu}$&$6.37 \pm 2.55$&$0.40 \pm 0.19$&$1.08 \pm 0.39$&$7.85 \pm 3.13$ \\
\end{tabular}
\end{ruledtabular}
\end{table}

\subsection{Nonleptonic decays}
Finally,  we want to evaluate the branching ratios for the
nonleptonic $D\to T~P (P=K, \pi)$  decays. For these decays, the
factorizable amplitude  has the expression \cite{HCheng2003}
\begin{eqnarray}
\mathcal{X}(D\to
T~P)&=&i\frac{G_{F}}{\sqrt{2}}\,V_{cq'}\,V_{uq_{_2}}^{*}\,f_{P}\,
\varepsilon^{*}_{\mu\nu}\,p^{\mu}\,p^{\nu}\Bigg[\mathcal{A}_{1}(m^{2}_{P})+(m^{2}_{D}-m^{2}_{P})\,\mathcal{A}_{2}(m^{2}_{P})
+m^{2}_{P}\,\mathcal{A}_{3}(m^{2}_{P})
\Bigg]\nonumber\\
&=&\varepsilon^{*}_{\mu\nu}\,p^{\mu}\,p^{\nu}\,\mathcal{M}(D\to
T~P),
\end{eqnarray}
where $q_{2}=s~(d)$ for $P=K~(\pi)$. Also,  $f_{P}$ is the $P$
meson decay constant, and $\mathcal{A}_{i}~~(i=1,2,3)$ is defined in
Eq. (\ref{ffdef}). The decay rate is given by
\begin{eqnarray}\label{nonld}
\Gamma(D\to T~P )=\frac{p_{c}^{5}}{12~\pi\,m_{T}^{2}}\left( \frac{m_{D}}{m_{T}}\right)^{2}\,
|\mathcal{M}(D\to T~P)|^{2},
\end{eqnarray}
where $p_{c}$ is the c.m. momentum of the tensor meson in the
rest frame of the $D$ meson. For estimating
the branching ratios of the nonleptonic $D\to T~P$ decays,
the values of $\mathcal{A}_{i}~(i=1, 2, 3)$ at $q^{2}=m_{P}^{2}$ are needed.
For $\pi $ and the $K$ meson, the masses are chosen in giga-electron-volts
as   $m_{\pi^{+}}=0.139 $ and $m_{K^{+}}=0.493$ \cite{pdg}.
The results are presented in Table \ref{ffmp}.
\begin{table}[th]
\caption{The values of $\mathcal{A}_{i} (i=1, 2, 3)$  for $D\to T~P (P=K, \pi)$ transition at $q^2=m^{2}_{P}$.} \label{ffmp}
\begin{ruledtabular}
\begin{tabular}{cccc}
$(D, T, P)$&$\mathcal{A}_{1}(m^{2}_{P})$& $\mathcal{A}_{2}(m^{2}_{P})$&  $\mathcal{A}_{3}(m^{2}_{P})$\\
\hline
$(D^{0}, a_{2}, K^{+})$&$-3.85 \pm 0.09$& $-0.88 \pm 0.24$&  $35.32 \pm 39.07$\\
$(D^{0}, K_{2}^{*}, \pi^{+})$&$-2.61 \pm 0.75$& $-1.16 \pm 0.30$&  $1.51 \pm 23.05$\\
$(D^{+}_{s}, f_{2}', \pi^{+})$&$-1.64\pm 0.31$& $-1.31 \pm 0.34$&  $3.41 \pm 66.80$\\
$(D^{+}_{s}, f_{2}', K^{+})$&$-1.39\pm 0.25$& $-1.21 \pm 0.29$&  $0.13 \pm 4.54$\\
$(D^{+}, f_{2}, \pi^{+})$&$-4.25 \pm 0.62$& $-0.72 \pm 0.18$&  $41.05 \pm 37.29$\\
\end{tabular}
\end{ruledtabular}
\end{table}
Inserting these values in Eq. (\ref{nonld}) and using
 $|V_{ud}|=(0.97 \pm 0.00)$, $|V_{us}|=(0.22 \pm  0.00)$,
$f_{\pi^{+}}=(130 \pm 0.26)~ \rm{MeV}$, and $f_{K^{+}}=(156 \pm 0.49)~ \rm{MeV}$,
the values for the branching ratio of nonleptonic decays are obtained as presented
 in Table \ref{nonleptonicBr}.
In comparison, the experimental values and IGSW results are also included in this table.
This table shows that for the $D^0\to a_{2}\,K^{+}$, $D^0\to K_{2}^{*}\,\pi^{+}$, and
$D^{+}\to f_{2}\,\pi^{+}$  cases our results for the branching ratios
are in good agrement with the experimental results.
\begin{table}[th]
\caption{Branching ratios for various $D\to T~P (P=K, \pi)$ decays.} \label{nonleptonicBr}
\begin{ruledtabular}
\begin{tabular}{cccc}
Decay&IGSW\cite{HCheng2003}&This work&Exp \cite{E791,BaBar020,CLEO2002}  \\
\hline
{Br}$(D^0\to a_{2}\,K^{+})\times 10^4$&$0.05 $&$7.06 \pm 1.44$&$7.0 \pm 4.3$ \\
{Br}$(D^0\to K_{2}^{*}\,\pi^{+})\times 10^3$&$0.10 $&$3.94 \pm 0.54$&$2.0^{ +1.3}_{-0.7} $ \\
{Br}$(D^{+}_{s}\to f_{2}'\,\pi^{+})\times 10^3$&$1.6$&$4.58\pm 0.42$& \\
{Br}$(D^{+}_{s}\to f_{2}'\,K^{+})\times 10^6$&$4.9$&$6.92 \pm 2.94$& \\
{Br}$(D^{+}\to f_{2}\,\pi^{+})\times 10^3$&$0.02$&$2.86 \pm 0.68$&$0.9 \pm 0.1$ \\
\end{tabular}
\end{ruledtabular}
\end{table}

In summary,  the $D \to T \ell \bar{\nu}_{\ell}$ decays in the LCSR
approach up to the twist-4 LCDAs of the $T$ tensor meson were
considered. Using  the transition form factors of the $D\to T $,
the semileptonic  branching ratios for $\ell=\mu$ and the
nonleptonic ones for $D\to T~P (P=K, \pi)$ decay were analyzed. For
the nonleptonic case, a comparison of the results for the branching
ratios with the IGSW approach and existing experimental results was
also made.

\clearpage
\appendix
\section{Form Factor Expressions }\label{app:form factors}
In this Appendix, the explicit expressions for  the  form factors of
$D \to T $ decays are given:
\begin{eqnarray}\label{eq3ffdef}
V(q^{2})&=&
\frac{\alpha_{1}}{2}{(m_{T}+ m_{D})}\,
\Bigg\{-f_{T} m_{c}\int_{u_0}^{1}du\,\frac{g_{a}^{(i)T}}{ 2 u^3 M^4} (2 u+3)~ e^{-s(u)}+
\frac{1}{4} f_{T}  \int_{u_0}^{1}du\,\frac{\Phi_{\perp}^{(i)T}}{ u^2 M^{2} } (2 u -7)~ e^{-s(u)}
-8 f_{T}^{\perp}m_{T}\nonumber\\
&&\times\int_{u_0}^{1}du\,\frac{h_{t}^{(iii)T}}{u^2 M^{4}} ~ e^{-s(u)}
-\frac{1}{8} f_{T}^{\perp}m_{T}\int_{u_0}^{1}du\,\frac{\bar{h}_{t}^{(iii)T}}{u^3 M^{4}} ~ (2 u+1)~ e^{-s(u)}
-\frac{1}{2} f_{T}^{\perp}m_{T}\int_{u_0}^{1}du\,\frac{\bar{h}_{3}^{(ii)T}}{u^2 M^{4}}~ (u-1)~ e^{-s(u)}
\nonumber\\
&-&\frac{16}{3}\,f_{T}^{\perp}m_{T}\int_{u_0}^{1}du\,\frac{{h}_{3}^{(ii)T}}{u^3 M^{4}}~ (u-1)~ e^{-s(u)}
\Bigg\},\nonumber\\
A_{2}(q^{2})&=&
\alpha_{1}\,{(m_{T}+ m_{D})}\,
\Bigg\{f_{T} m_{c}\int_{u_0}^{1} du \frac{\left[9\,\Phi_{\|}^{(ii)T}-g_{v}^{(ii)T}+
2\frac{m_{T}^{2}}{M^2}\bar{g}_{3}^{(iii) T}(u)\right]}{u^3\,M^{4}}\,~ e^{-s(u)}
-\frac{17 f_{T}^{\perp}}{8 m_{T}} \int_{u_0}^{1}du\frac{\Phi_{\perp}^{(i)T}}{u^2\,M^{2}}~ e^{-s(u)}\nonumber\\
&+& 2~f_{T}^{\perp}\,m_{T}\int_{u_0}^{1}du \frac{h_{t}^{(iii)T}}{u^3 M^{4} }\Bigg[
39 +  4 u-24\, \frac{\delta_{2}(u)}{M^{2}}+ 48~ (1-\frac{\delta_{1}(u)}{3 M^{2}})
\Bigg] ~ e^{-s(u)}
+\frac{1}{2} f_{T}^{\perp}m_{T}\int_{u_0}^{1}du\, \frac{h_{s}^{(i) T}}{u^{2}  M^{2}}
~ e^{-s(u)}\nonumber\\
&-&\frac{1}{3}\,f_{T}\,m_{T}^{2}\,m_{c}\int_{u_0}^{1}du\, \frac{g_{3}^{(iii) T}}{u^{3}  M^{6}}
~ e^{-s(u)}
+2\,f_{T}^{\perp}m_{T}\int_{u_0}^{1}du\, \frac{h_{3}^{(ii) T}\,(2-u)}{u^{2}  M^{4}}
~ e^{-s(u)}
\Bigg\},\nonumber\\
A_{0}(q^{2})&=&A_{3}(q^{2})-\frac{\alpha_{1}\,q^2}{2\,m_{T}}
\Bigg\{f_{T} \, m_{c}\int_{u_0}^{1} du\frac{ \left[-9\,\Phi_{\|}^{(ii)T}+g_{v}^{(ii)T}+
+(4- 2 u)\frac{m_{T}^{2}}{M^2}\,\bar{g}_{3}^{(iii) T}(u)\right]}{u^3\,M^{4}}\,~ e^{-s(u)}
+\frac{17 f_{T}^{\perp}}{8 m_{T}}\nonumber\\
&&\times \int_{u_0}^{1}du\frac{\Phi_{\perp}^{(i)T}}{u^2\,M^{2}}~ e^{-s(u)}
+ 2~f_{T}^{\perp}\,m_{T}\int_{u_0}^{1}du \frac{h_{t}^{(iii)T}}{u^3 M^{4} }\Bigg[
- 31 -  4  u + 24\, \frac{\delta_{2}(u)}{ M^{2}}+ 48~ (4- 2 u)
\Bigg] ~ e^{-s(u)}
\nonumber\\
&+& \frac{1}{4} f_{T}^{\perp}m_{T}\int_{u_0}^{1}du\, \frac{h_{s}^{(i) T}}{u^{3}  M^{4}} (2-u) \Bigg[
1-\frac{\delta_{1}(u)}{3 M^{2}}\Bigg]
~ e^{-s(u)}
-\frac{1}{6}\,f_{T}\,m_{T}^{2}\,m_{c}\int_{u_0}^{1}du\, \frac{g_{3}^{(iii) T}\,(2-u)}{u^{4}  M^{6}}
~ e^{-s(u)}
\nonumber\\
&+&2\,f_{T}^{\perp}m_{T}\int_{u_0}^{1}du\, \frac{h_{3}^{(ii)
T}\,(2-u)^2}{u^{3}  M^{4}} ~ e^{-s(u)} \Bigg\}.
\end{eqnarray}


\clearpage

\begin{thebibliography}{II}

\bibitem{Khodjamirian}
A. Khodjamirian, R. Ruckl, S. Weinzierl, C. Winhart, and O. I. Yakovlev, Phys.Rev. D {\bf 62},
114002 (2000).

\bibitem{BallD}
P. Ball, Phys. Lett. B {\bf 641}, 50 (2006).


\bibitem{Abada}
A. Abada $et~ al$. (SPQcdR Collaboration), Nucl.  Phys. B, Proc. Suppl. {\bf 119}, 625 (2003).

\bibitem{Aubin}
C. Aubin $et~ al$. (Fermilab Lattice Collaboration), Phys. Rev. Lett. {\bf 94}, 011601 (2005).

\bibitem{Bernard}
C. Bernard $et~ al$., Phys. Rev. D {\bf 80}, 034026 (2009).
\bibitem{WangD}
W. Y. Wang, Y. L. Wu, and M. Zhong, Phys. Rev. D {\bf 67}, 014024 (2003).

\bibitem{Ignacio}
I. Bediaga and M. Nielsen, Phys. Rev. D {\bf 68}, 036001 (2003).


\bibitem{Aliev}
T. M. Aliev, V. L. Eletsky, and Ya. I. Kogan, Sov. J. Nucl. Phys.
{\bf 40}, 527 (1984).

\bibitem{Ball2}
P. Ball, V. M. Braun, and H. G. Dosch, Phys. Rev. D {\bf 44}, 3567
(1991).

\bibitem{Ball3}
P. Ball, Phys. Rev. D {\bf 48}, 3190 (1993).

\bibitem{Ovchinnikov}
A. A. Ovchinnikov and V. A. Slobodenyuk, Z. Phys. C {\bf 44}, 433
(1989).

\bibitem{Baier90}
V. N. Baier and A. Grozin, Z. Phys. C {\bf 47}, 669 (1990).


\bibitem{Dong}
D. S. Du, J. W. Li, and M. Z. Yang, Eur. Phys. J. C {\bf 37}, 137
(2004).

\bibitem{Mao}
M. Z. Yang, Phys. Rev. D {\bf 73}, 034027 (2006).

\bibitem{Khosravi}
R. Khosravi, K. Azizi, and N. Ghahramany, Phys. Rev. D {\bf 79},
036004 (2009).

\bibitem{Zuo}
Y. Zuo $et~ al$, Int. J. Mod. Phys. A { \bf 31}, 1650116 (2016).

\bibitem{Isgur1989}
N. Isgur, D. Scora, B. Grinstein, and M.B. Wise, Phys. Rev. D {\bf 39}, 799 (1989).

\bibitem{Scora1995}
D. Scora and N. Isgur, Phys. Rev. D {\bf 52}, 2783 (1995).

\bibitem{Hagiwara2002}
K. Hagiwara $et~ al$. (Particle Data Group) Phys. Rev. D {\bf 66},
010001 (2002).

\bibitem{Li2001}
D. M. Li, H. Yu, and Q. X. Shen, J. Phys. G {\bf 27}, 807 (2001).

\bibitem{Cheng2010}
H. Y. Cheng, Y. Koike, and K. C. Yang, Phys. Rev. D {\bf 82}, 054019 (2010).

\bibitem{Hatanaka2009}
H. Hatanaka and K. C. Yang, Phys. Rev. D {\bf 79}, 114008 (2009).

\bibitem{Hatanaka2010}
 H. Hatanaka and K. C. Yang, Eur. Phys. J. C {\bf 67}, 149 (2010).

\bibitem{Wang2011}
W. Wang, Phys. Rev. D {\bf 83}, 014008 (2011).

\bibitem{Yangt2011}
K. Yang, Phys. Lett. B {\bf 695}, 444 (2011).

\bibitem{Chengt2010}
H. Cheng, Y. Koike, and K. Yang, Phys. Rev. D {\bf 82}, 054019 (2010).

\bibitem{Ball99}
P. Ball and V. M. Braun, Nucl. Phys.  {\bf B 543}, 201 (1999).

\bibitem{pdg}
K. A. Olive $et~ al$. (Particle Data Group), Chin. Phys. C  {\bf 38}, 090001 (2014) and (2015) update.

\bibitem{Huang2004}
M. Q. Huang, Phys. Rev. C {\bf 69}, 114015 (2004).

\bibitem{Mutuk}
H. Mutuk,   Adv. High Energy Phys. {\bf 2018}, 8095653 (2018).

\bibitem{Chengb}
H.Y. Cheng, C.K. Chua,  and C.W. Hwang, Phys. Rev. D {\bf 69}, 074025 (2004).

\bibitem{HCheng2003}
H. Y.  Cheng, Phy. Rev.  D {\bf 68}, 014015 (2003).

\bibitem{E791}
E. M. Aitala $et~ al$. (E791 Collaboration), Phys. Rev. Lett. {\bf 89},
121801 (2002).

\bibitem{BaBar020}
 B. Aubert $et~ al$. (BABAr Collaboration), arxiv: hep-ex /0207089.

\bibitem{CLEO2002}
H. Muramatsu $et~ al$. (CLEO Collaboration), Phys. Rev. Lett. {\bf 89}, 251802 (2002).




\end{thebibliography}
\end{document}